\newcommand{\Ind}{\mathbb{I}} 

\newcommand{\mbs}[1]{\mathbf{#1}}

\documentclass[preprint,12pt,authoryear]{elsarticle} 

\usepackage{amssymb}
\usepackage[margin=1in]{geometry}
\usepackage{amsmath}
\usepackage{booktabs}

\usepackage{algorithm}
\usepackage{xcolor}
\usepackage{algpseudocode}
\usepackage{adjustbox}

\usepackage[colorlinks=false]{hyperref}
\usepackage{setspace}
\journal{Spatial Statistics}

\begin{document}

\begin{frontmatter}



\author[GMU]{Isaac Amouzou} 
\author[GMU]{Seiyon Ben Lee}

\affiliation[gmustat]{
    organization={Department of Statistics, George Mason University},
    addressline={4400 University Dr},
    city={Fairfax},
    state={VA},
    postcode={22030},
    country={USA}
}



\title{A Cubing Strategy for Identifying Stable Hyperparameter Regions for Uncertainty Quantification in Spatial Deep Learning}
\begin{abstract}
Spatially referenced datasets have become increasingly prevalent across 
many fields, largely driven by advances in data collection methods such as satellite remote sensing. In many applications, predictions at unobserved locations are accompanied by reliable uncertainty estimates. While deep learning methods provide both scalable and accurate models for spatial predictions, there remains no clear consensus for addressing uncertainty quantification in spatial deep learning. Monte Carlo (MC) dropout has become a popular approach for uncertainty quantification, yet existing implementations typically focus on tuning the dropout rate while fixing other influential hyperparameters, such as weight decay and the predictive standard deviation multiplier, often through ad-hoc or manual tuning. We propose a cubing-based diagnostic framework that recursively partitions the hyperparameter space to identify stable regions where MC dropout yields well-calibrated predictive intervals. The approach evaluates hyperparameter regions using scoring rules relative to a statistical baseline model, which serves as a calibration anchor. Through a simulation study spanning multiple spatial dependence regimes as well as a large remotely-sensed land surface temperature dataset, we demonstrate that our approach produces competitive or superior predictive intervals compared to the baseline model. Our methodology provides practitioners with a systematic procedure for incorporating uncertainty quantification into spatial deep learning models.

\end{abstract}


\end{frontmatter}



\section{Introduction}
\label{sec1}

Spatially referenced data are prevalent in a wide range of scientific fields, including environmental science \citep{ramcharan2018soil}, mining \citep{senanayake2023spatial}, criminology \citep{quick2018crime}, economics \citep{redding2017quantitative}, hydrology \citep{zhou2020geo}, public health \citep{rushton2003public}, and genetics \citep{wagner2013conceptual}. Recent advances in remote sensing technologies and other data acquisition methods have enabled the collection of high-dimensional spatial datasets that contain tens of thousands of locations. Such large datasets pose significant computational challenges for statistical modeling, particularly for prediction and uncertainty quantification at unobserved locations.

Traditional spatial statistical models, such as the spatial linear mixed model (SLMM) \citep{cressie1993statistics, banerjee2014hierarchical}, capture spatial dependence through spatially correlated random effects modeled as a latent Gaussian process with a spatial covariance function. However, SLMMs become computationally prohibitive when the number of observation locations is moderately large due to large matrix operations \citep{haran2003accelerating, cressie2008fixed, wikle2019spatio}. Specifically, evaluating Gaussian likelihood requires a Cholesky factorization of a dense $n\times n$ covariance matrix that has computational costs on the order of $\mathcal{O}(n^3)$ and storage costs $\mathcal{O}(n^2)$. Basis representation approaches \citep{cressie2022basis} address this issue by approximating the latent spatial random effect as a linear combination of spatial basis functions; thus reducing computational cost by bypassing large matrix operations \citep{Banerjee2008Gaussian,higdon1998process}, reducing the dimension of spatial random effects\citep{Guan_Haran_2018,sglmm}, and enabling sparse matrix computations by using localized basis functions \citep{katzfuss2017multi,nychka2015multiresolution,benedetti2022identifying}. Common choices for spatial basis functions include radial basis functions \citep{cressie2008frk}, predictive process basis functions \citep{Banerjee2008Gaussian}, Moran's basis functions for spatial eigenvector filtering \citep{sglmm,Griffith2003Spatial-Autocor}, piecewise linear basis functions defined on triangulated meshes \citep{lindgren2011explicit}, and convolution kernel representations \citep{higdon1998process}. While these approaches reduce the computational cost to roughly 
$\mathcal{O}(nm+m^3)$, where $m$ is the number of basis functions, they can still be computationally demanding for large datasets.

The use of deep learning methods to model spatial data has grown substantially in recent years, and we refer the reader to the review by  \citet{wikle2023statistical} for a comprehensive overview. Several works have enhanced classical geostatistical methods using deep learning. For example, 
\citet{burmeister2023deep} replace the parametric variogram in kriging with a deep neural network to learn anisotropic spatial correlations. \citet{chen2024deepkriging} introduce DeepKriging, which incorporates spatial dependence by adding spatial basis functions of the coordinates as inputs to a deep neural network. This formulation links kriging with neural network models, while deeper nonlinear architectures allow more flexible modeling of nonstationary and non-Gaussian spatial processes. \citet{zhan2024nngls} propose NN-GLS, which embeds a neural network within a Gaussian process spatial model and estimates parameters using a generalized least squares loss that explicitly accounts for spatial covariance. When using a Nearest Neighbor Gaussian Process covariance structure \citep{datta2016hierarchical}, the NN-GLS reduces to a graph neural network, enabling scalable inference for large spatial datasets. Despite these advances, providing reliable uncertainty quantification alongside deep learning spatial predictions remains an open challenge.

In deep learning, Monte Carlo (MC) dropout has been used for uncertainty quantification by averaging predictions obtained from randomly thinned versions of the network \citep{gal2016dropout}. To implement MC dropout, key hyperparameters must be tuned, such as the dropout rate and weight decay, and one must decide how to handle different sources of uncertainty (e.g., epistemic and aleatoric). In practice, practitioners often resort to ad-hoc or manual tuning \citep{kirkwood2022bayesian} or grid-search strategies \citep{gal2017concrete}. For example \citet{wang2025spatial} incorporate radial basis functions within a convolutional neural network to represent spatial data as images, while \citet{kirkwood2022bayesian} take a similar approach. In both cases, MC dropout is employed for uncertainty quantification, though key tuning parameters such as the dropout rate and weight decay are fixed or manually tuned rather than learned from data. Poor choices of the dropout rate can lead to unreliable uncertainty estimates \citep{wu2021quantifying,zhu2017deep,verdoja2020notes}, and misspecified weight decay may lead to overconfident predictions \citep{liu2021peril,de2025deep}. Despite the widespread use of MC dropout, there is currently no systematic approach for determining hyperparameter regions that produce well-calibrated prediction uncertainties. 

To address this limitation, we propose a cubing-based diagnostic framework that identifies stable hyperparameter regions using scoring rules. The proposed approach identifies stable subregions that emulate a baseline model as a reference. We demonstrate our approach through a comprehensive simulation study that examines many different spatial dependence settings, and we apply our method to a high-dimensional dataset of remotely sensed land surface temperatures. In practice, the cubing strategy can be used during model development and then the results can potentially be reused or transferred across datasets. To our knowledge, this is the first systematic framework for identifying stable hyperparameter regions that yield calibrated predictive uncertainty in spatial deep learning models.

In Section~\ref{SLMM}, we provide an overview of traditional spatial regression models for Gaussian data and their basis representation extensions. Section~\ref{sec:NB_SLMM} reviews neural-SLMMs, a neural network framework used to approximate the spatial linear mixed model, and discusses approaches for uncertainty quantification. The proposed cubing strategy for identifying stable hyperparameter subregions is presented in Section~\ref{sec:Cubing}, along with implementation details and the corresponding algorithm. We evaluate the proposed cubing approach through an extensive simulation study in Section~\ref{simstudy} and a high-dimensional satellite imagery application in Section~\ref{sec:application}. Finally, Section~\ref{sec:discussion} concludes with a discussion and directions for future research.

\section{Hierarchical Spatial Linear Mixed Models (SLMMs)}
\label{SLMM}

Let $\mathbf{s} \in \mathcal{D} \subset \mathbb{R}^2$, where $\mathcal{D}$ is a spatial domain in two-dimensional Euclidean space. For observed locations $\mbs{s_i}$, let $\{Z(\mathbf{s})\}$ denote a spatial stochastic process. 
At location $\mbs{s_i}$, we model
\[
z(\mathbf{s}_i) = \mathbf{x}(\mathbf{s}_i)^\top \boldsymbol{\beta} 
+ \omega(\mathbf{s}_i) 
+ \epsilon(\mathbf{s}_i), \quad i=1,\ldots,n,
\]
where $\mathbf{x}(\mathbf{s}_i)$ is a $p$-dimensional vector of covariates with corresponding regression coefficients $\boldsymbol{\beta} \in \mathbb{R}^p$, $\{\omega(\mathbf{s}_i)\}$ represents a latent spatial process that induces spatial dependence, and $\epsilon(\mathbf{s}_i)$ denotes the iid measurement error where $\epsilon(\mathbf{s}_i) \sim \mathcal{N}(0, \tau^2)$. 

Over a continuous spatial domain, $\{\omega(\mathbf{s})\}$ is typically modeled as a Gaussian process with covariance function  $C_{\boldsymbol{\theta}}(\mathbf{s}_i,\mathbf{s}_j)$, parameterized by $\boldsymbol{\theta}$. The covariance function generates a positive definite covariance matrix $\mathbf{C}_{\boldsymbol{\theta}}$ over observed locations, where the covariance $C_{\boldsymbol{\theta}}(\mathbf{s}_i,\mathbf{s}_j)$ generally decreases as the two points $\mathbf{s}_i$ and $\mathbf{s}_j$ grow further apart. The Mat\'ern class \citep{williams2006gaussian,stein1999interpolation} is a flexible class of stationary and isotropic covariance functions defined as:
 \begin{equation}\label{eq:Matern}
     C_{\theta}(\mathbf{s}_i,\mathbf{s}_j)=\sigma^2 \frac{2^{1-\nu}}{\Gamma(\nu)} \big(\sqrt{2\nu} \frac{d}{\rho}\big)^\nu K_\nu\big(\sqrt{2\nu} \frac{d}{\rho}\big),
 \end{equation}
where the parameter vector is $\pmb{\theta}=(\nu, \sigma^2, \rho)^\top$ with smoothness $\nu>0$ that controls the mean-square differentiability of the process \citep{stein1999interpolation}, marginal variance $\sigma^2>0$, and the range $\rho$, which controls how fast correlation will decay with respect to the pairwise Euclidean distance $d=||\mathbf{s}_i-\mathbf{s}_j||$.

To simplify notation, we define vectors $\mathbf{Z} = \{z(s_i)\}^N_{i=1}$ and $\boldsymbol{\omega} = \{\boldsymbol{\omega}(s_i)\}^N_{i=1}$ for location indices $1,...,N$. For a finite set of locations, the observation vector is modeled as:
\begin{equation}
 \mathbf{Z}\sim \mathcal{N}_N(\mathbf{X}\boldsymbol{\beta}+\boldsymbol{\omega}, \tau^2 \mathbf{I}_N)   
\end{equation}
where $\mathbf{X}$ represents the design matrix and $\mathbf{I}_N$ is the N-dimensional identity matrix. To complete the Bayesian hierarchical model, we place prior distributions on parameters $\boldsymbol{\beta}$ and covariance parameters $\boldsymbol{\theta}$ and $\tau^2$. The resulting Bayesian hierarchical model is:
\begin{equation}
\begin{aligned}
\text{Data model:}\quad &
\mathbf{Z} \mid \boldsymbol{\beta}, \boldsymbol{\omega}, \tau^2 
\sim \mathcal{N}_N(\mathbf{X}\boldsymbol{\beta} + \boldsymbol{\omega}, \tau^2 \mathbf{I}_N), \\
\text{Process model:}\quad &
\boldsymbol{\omega} \mid \boldsymbol{\theta} 
\sim \mathcal{N}_N(\mathbf{0}, \boldsymbol{\Sigma}_{\boldsymbol{\theta}}), \\
\text{Parameter model:}\quad &
\boldsymbol{\beta} \sim p(\boldsymbol{\beta}),\quad \tau^2 \sim p(\tau^2),\quad \boldsymbol{\theta} \sim p(\boldsymbol{\theta})
\end{aligned}
\label{eq:SLMM}
\end{equation}
\noindent where $\boldsymbol{\Sigma}_\theta$ is the resulting $N\times N$ positive definite covariance matrix acquired from function $C_{\theta}(\cdot, \cdot)$, and $p(\cdot)$ denotes the chosen prior distributions.  

Upon integration of the spatial random effects $\boldsymbol{\omega}$, we arrive at an equivalent formulation:
\begin{equation}\label{eq:model_collapsed}
 \mathbf{Z}\sim \mathcal{N}_N(\mathbf{X}\boldsymbol{\beta}, \boldsymbol{\Sigma}_\theta+ \tau^2 \mathbf{I}_N)   
\end{equation}
with the corresponding Bayesian hierarchical model:
\begin{equation}
\begin{aligned}
\text{Data model:}\quad &
\mathbf{Z} \mid \boldsymbol{\beta}, \tau^2 
\sim \mathcal{N}_N(\mathbf{X}\boldsymbol{\beta}, \boldsymbol{\Sigma}_\theta+ \tau^2 \mathbf{I}_N), \\
\text{Parameter model:}\quad &
\boldsymbol{\beta} \sim p(\boldsymbol{\beta}),\quad \tau^2 \sim p(\tau^2),\quad \boldsymbol{\theta} \sim p(\boldsymbol{\theta})
\end{aligned}
\end{equation}

The primary computational bottleneck arises from evaluating the multivariate normal likelihood in~\eqref{eq:model_collapsed}, which requires computing the log-determinant and quadratic form involving $\boldsymbol{\Sigma}_\theta + \tau^2 \mathbf{I}_N$. These operations rely on a Cholesky factorization of the $N \times N$ covariance matrix and incur $\mathcal{O}(N^3)$ computational complexity and $\mathcal{O}(N^2)$ memory requirements.



\subsubsection{Basis Representation for SLMMs}\label{subsec:basis_SLMM}
In spatial datasets, basis expansions have been widely used to approximate latent spatial random effects \citep{cressie2008fixed,cressie2022basis}. Specifically, the latent spatial process \(\boldsymbol{\omega}\) is approximated by a linear combination of \(m\) spatial basis functions \(\{\phi_j(\mathbf{s})\}_{j=1}^m\), so that $\boldsymbol{\omega} \approx \boldsymbol{\Phi}\boldsymbol{\gamma}$
where \(\boldsymbol{\Phi} = [\boldsymbol{\phi}_1,\boldsymbol{\phi}_2,\dots,\boldsymbol{\phi}_m]\) is the \(N\times m\) basis matrix, \(\boldsymbol{\phi}_j \in \mathbb{R}^N\) contains the evaluations of the \(j\)th basis function at the observed locations, and \(\boldsymbol{\gamma}\in\mathbb{R}^m\) denotes the corresponding basis coefficients. The resulting basis-representation spatial linear mixed model is
\begin{equation}\label{eq:basis_slmmm}
z(\mathbf{s}_i)=\mathbf{x}(\mathbf{s}_i)^\top \boldsymbol{\beta}
+\boldsymbol{\phi}(\mathbf{s}_i)^\top \boldsymbol{\gamma}
+\epsilon(\mathbf{s}_i), \qquad i=1,\dots,N,
\end{equation}
where \(\epsilon(\mathbf{s}_i)\stackrel{\mathrm{iid}}{\sim}\mathcal{N}(0,\tilde{\tau}^2)\).

A wide range of spatial basis functions have been proposed for approximating spatial random effects, including radial basis functions such as bi-square bases, empirical orthogonal functions (EOFs), wavelet bases, and multiresolution bases \citep{cressie2008fixed,nychka2002multiresolution,cressie2015statistics,nychka2015multiresolution,katzfuss2017multi}.

Let \(\tilde{\mathbf X}=[\mathbf X,\boldsymbol{\Phi}]\) denote the augmented design matrix obtained by concatenating the covariates and basis functions, and let \(\tilde{\boldsymbol{\beta}}=(\boldsymbol{\beta}^\top,\boldsymbol{\gamma}^\top)^\top\) denote the corresponding regression coefficients. Then the model can be written compactly as
\[
\mathbf Z = \tilde{\mathbf X}\tilde{\boldsymbol{\beta}} + \tilde{\boldsymbol{\varepsilon}},
\qquad
\tilde{\boldsymbol{\varepsilon}}\sim \mathcal{N}_N(\mathbf 0,\tilde{\tau}^2 \mathbf I_N),
\]
where \(\tilde{\tau}^2\) denotes the microscale variance. After specifying prior distributions for \(\tilde{\boldsymbol{\beta}}\) and \(\tilde{\tau}^2\), the Bayesian hierarchical model is given by
\begin{equation}
\begin{aligned}
\text{Data model:}\quad &
\mathbf Z \mid \tilde{\boldsymbol{\beta}}, \tilde{\tau}^2
\sim \mathcal{N}_N(\tilde{\mathbf X}\tilde{\boldsymbol{\beta}}, \tilde{\tau}^2 \mathbf I_N), \\
\text{Parameter model:}\quad &
\tilde{\boldsymbol{\beta}} \sim p(\tilde{\boldsymbol{\beta}}), \qquad
\tilde{\tau}^2 \sim p(\tilde{\tau}^2),
\end{aligned}
\end{equation}
\noindent where $p(\cdot)$ denotes the prior distributions for $\tilde{\boldsymbol{\beta}}$ and $\tau^2$. In contrast to the spatial linear model in~\ref{eq:model_collapsed}, the basis representation approach requires $\mathcal{O}(Nm)$ storage and computational cost, which can still be computationally prohibitive for a large number of locations or basis functions. 

\section{Neural Basis Spatial Linear Mixed Models (NB-SLMM)}
\label{sec:NB_SLMM}
This section provides an overview of the neural network framework used to approximate the spatial linear mixed model described in~\eqref{eq:basis_slmmm}. Let $(\mathbf{x}_i,Z(\mathbf{s}_i))$ denote observations from a spatial process at locations $\mbs{s_i}$, where $\mathbf{x}_i\in\mathbb{R}^{p}$ represents input features and $z(\mathbf{s}_i)\in\mathbb{R}$ denotes the observed response. The input features include both covariates and spatial basis functions employed in~\eqref{eq:basis_slmmm}. Specifically, we define the feature vector $\mathbf{x}_\phi(\mathbf{s}_i)=\big(\mathbf{x}(\mathbf{s}_i)^\top,\boldsymbol{\phi}(\mathbf{s}_i)^\top\big)^\top$, where $\mathbf{x}(\mathbf{s}_i)$ denotes covariates and $\boldsymbol{\phi}(\mathbf{s}_i)$ represents spatial basis functions. Rather than modeling the components of the spatial linear mixed model in~\eqref{eq:basis_slmmm} explicitly, we approximate the conditional mean function using a feedforward neural network. $\mathbb{E}\!\left[z(\mathbf{s}_i)\mid\mathbf{x}_\phi(\mathbf{s}_i)\right]
\approx f(\mathbf{x}_\phi(\mathbf{s}_i);\boldsymbol{\psi})$,
where $f(\cdot;\boldsymbol{\psi})$ denotes a neural network with unknown parameters $\boldsymbol{\psi}$. The network implicitly captures both the fixed effects $\mathbf{x}(\mathbf{s}_i)^\top\boldsymbol{\beta}$ and the spatial component $\boldsymbol{\phi}(\mathbf{s}_i)\boldsymbol{\gamma}$ in a flexible nonlinear model. 

\subsection{Feedforward Neural Networks}

Let $p$ refer to the number of covariates and $K$ refer to the number of latent variables, a feedforward neural network (FNN) with $L$ layers defines a mapping $f:\mathbb{R}^{p+K}\rightarrow\mathbb{R}$ through a sequence of affine transformations and nonlinear activation functions. Let $\mathbf{a}^{(0)}=\mathbf{x}_\phi(\mathbf{s})$ denote the input layer. For $\ell=1,\dots,L$, the network computes
\[
\mathbf{z}^{(\ell)}=\mathbf{W}^{(\ell)}\mathbf{a}^{(\ell-1)}+\mathbf{b}^{(\ell)},\qquad
\mathbf{a}^{(\ell)}=\sigma^{(\ell)}\!\big(\mathbf{z}^{(\ell)}\big),
\]
where $\mathbf{W}^{(\ell)}$ and $\mathbf{b}^{(\ell)}$ denote weight matrices and bias vectors, respectively, and $\sigma^{(\ell)}(\cdot)$ represents a nonlinear activation function such as the rectified linear unit (ReLU) or hyperbolic tangent. The collection of weights and biases across all layers defines the network parameters $\boldsymbol{\psi}=\{\mathbf{W}^{(\ell)},\mathbf{b}^{(\ell)}\}_{\ell=1}^{L}.$

Nonlinear activation functions $\sigma^{(\ell)}(\cdot)$ enable neural networks to approximate complex functional relationships between inputs and outputs. In particular, the universal approximation theorem states that feedforward neural networks with sufficiently many hidden units can approximate any continuous function on a compact domain arbitrarily well \citep{hornik1989multilayer}. For regression problems, the output layer typically uses an identity activation function so that the predicted response is given by $
\hat{z}(\mathbf{s}_i)=f(\mathbf{x}_\phi(\mathbf{s}_i);\boldsymbol{\psi})
$. The network parameters are estimated by minimizing an empirical loss function over the training data (e.g., mean squared error for regression), which is typically optimized using stochastic gradient-based methods.


\subsection{Monte Carlo Dropout}\label{s:MC_Dropout}
To mitigate overfitting, dropout regularization \citep{srivastava2014dropout} is applied during training, where subsets of neurons are randomly deactivated at each stochastic gradient update. This mechanism prevents the neural network from heavily relying on specific features or hidden units, resulting in more robust model training. For network layer $\ell$, a random mask $\mathbf{m}^{(\ell)}$ is applied to the activations such that, $\tilde{\mathbf{a}}^{(\ell)} = \mathbf{m}^{(\ell)} \odot \mathbf{a}^{(\ell)},$ where $\odot$ denotes elementwise multiplication and each component of $\mathbf{m}^{(\ell)}$ is independently sampled from a Bernoulli distribution with probability $1-p$, where $p$ is the dropout rate. During training, dropout randomly removes neurons and their associated connections with probability $p$ at each stochastic gradient update. The remaining activations are scaled by $1/(1-p)$, which keeps the expected activations consistent across layers (i.e., inverted dropout).  This produces a different ``thinned'' network at each iteration, yet all subnetworks share the same underlying weights $\mathbf{W}^{(\ell)}$ and  biases $\mathbf{b}^{(\ell)}$. 

At prediction time, dropout is disabled and the full network is used without additional scaling. Consequently, the resulting prediction
$\hat{z}(\mathbf{s}) = f(\mathbf{x}_\phi(\mathbf{s});\hat{\boldsymbol{\psi}})$ is an approximation of the average prediction of the ensemble of subnetworks trained through dropout.

\paragraph{MC Dropout Predictive Draws and Prediction Intervals}

Let $\boldsymbol{\psi}^{(t)}$ denote the effective network parameters induced by the dropout mask on pass $t$. MC Dropout can be interpreted as a variational Bayesian approximation where dropout induces a Bernoulli distribution over weights, yielding an approximate posterior over the network weights. Within MC dropout, a forward pass represents a single evaluation of the neural network for a given input, where a randomly sampled dropout mask produces one stochastic realization of the prediction. For a new spatial location $\mathbf{s}_\ast$ with input features $\mathbf{x}_\phi(\mathbf{s}_\ast)$, the corresponding prediction is:
\[
\hat{y}_\ast^{(t)} = f(x_\ast;\boldsymbol{\psi}^{(t)}), \qquad t=1,\dots,T.
\]
\noindent Stochastic forward passes yield Monte Carlo samples from the predictive distribution, with
\[
\hat{\mu}(x_\ast)=\frac{1}{T}\sum_{t=1}^T \hat{y}_\ast^{(t)}, \qquad
\hat{\sigma}_{\text{epi}}^2(x_\ast)=\frac{1}{T}\sum_{t=1}^T\big(\hat{y}_\ast^{(t)}-\hat{\mu}(x_\ast)\big)^2,
\]
where the variance $\hat{\sigma}_{\text{epi}}^2(x_\ast)$ captures epistemic uncertainty from limited data. Epistemic uncertainty refers to uncertainty arising from limited data and model misspecification, reflecting a lack of knowledge about the true underlying process and, generally, reducible with additional information.

\subsection{Tuning Parameters}
Key hyperparameters influencing uncertainty quantification in MC dropout include: (i) the dropout rate $p$; (ii) weight decay $\lambda$; (iii) specification of the sources of uncertainty; and (iv) the chosen multiplier $k$ for the predictive standard deviation $\sigma_{tot}$.

\paragraph{MC Dropout Probability}
The dropout rate $p\in[0,1]$ influences the variability of the sampled subnetworks; hence, it is an important hyperparameter affecting epistemic uncertainty. Previous work has shown that fixing this parameter heuristically may lead to unstable uncertainty estimates \citep{wu2021quantifying,zhu2017deep}. 

\paragraph{Weight decay (L2 regularization)}
To improve generalization and control model complexity, a quadratic penalty on the network weights is added \citep{bishop2006pattern,krogh1991simple} with weight decay parameter $\lambda$:
\[
\min_{\boldsymbol{\psi}}\ \frac{1}{n}\sum_{i=1}^n 
\mathcal{L}\!\big(z(\mathbf{s}_i),f(\mathbf{x}_\phi(\mathbf{s}_i);\boldsymbol{\psi})\big)
\;+\; \frac{\lambda}{2}\sum_{\ell=1}^{L}\big\|\mathbf{W}^{(\ell)}\big\|_F^{2}.
\]

The parameter $\lambda$ shrinks weights toward zero by penalizing large parameter values, analogous to ridge regularization in linear models \citep{bishop2006pattern}. 


\paragraph{Sources of Uncertainty (Epistemic vs. Aleatoric)}
In regression problems, the total predictive uncertainty includes epistemic and aleatoric components. The aleatoric part represents irreducible noise such as measurement error or inherent variability in the data. Let $\hat{\sigma}_{\text{tot}}^2(x_\ast)
=
\hat{\sigma}_{\text{epi}}^2(x_\ast)
+
\hat{\sigma}_{\text{ale}}^2(x_\ast)$
denote the total predictive variance. In this study, we consider three approaches for accounting for the aleatoric variance component.

\begin{enumerate}

\item \textbf{Epistemic uncertainty only (EU):}
Predictive uncertainty arises solely from epistemic sources (see \citep{lops2025evaluating, amini2018spatial}),
\[
\hat{\sigma}_{\text{tot}}^2(x_\ast)
=
\hat{\sigma}_{\text{epi}}^2(x_\ast).
\]

\item \textbf{Fixed aleatoric variance (FA):}
A constant observation noise variance $\tau^{-1}$ is added to the epistemic component \citep{gal2016dropout},
\[
\hat{\sigma}_{\text{tot}}^2(x_\ast)
=
\hat{\sigma}_{\text{epi}}^2(x_\ast)
+
\tau^{-1}.
\]
The precision parameter $\tau$ is linked to the training hyperparameters through
$\tau = \frac{p\,\ell^{2}}{2N\lambda}$, where $p$ denotes the dropout keep probability, $\ell$ is a prior length scale, $N$ is the training sample size, and $\lambda$ is the weight decay parameter.

\item \textbf{Learned aleatoric variance (LA):}
Aleatoric uncertainty can be modeled directly by allowing the neural network to jointly predict the conditional mean and variance using a two-headed architecture \citep{kendall2017uncertainties}. Here, the network outputs the predicted mean and aleatoric variance $[\hat{y}_i,\hat{\sigma}_{a,i}^2]$, corresponding to a Gaussian likelihood with input-dependent variance. The model is trained by minimizing the Gaussian negative log-likelihood
\[
\frac{1}{n}\sum_{i=1}^{n}
\left[
\frac{(y_i-\hat{y}_i)^2}{2\hat{\sigma}_{a,i}^2}
+
\frac{1}{2}\log(\hat{\sigma}_{a,i}^2)
\right],
\]
where $\hat{\sigma}_{a,i}^2$ represents the predicted aleatoric variance for observation $i$.

\end{enumerate}

\noindent Prediction intervals are constructed using the estimated predictive distribution $\hat{\mu}(x_\ast) \pm k\,\hat{\sigma}_{\text{tot}}(x_\ast)$ where $k$ is chosen according to the desired confidence level (e.g., $k=1.96$ for approximate $95\%$ intervals under normality assumptions).


\paragraph{Predictive Variance Multiplier}
The assumed predictive variance multiplier, $k$, where $k$ is applied for $k\hat{\sigma}_{tot}$ ($k$ standard deviations) is another hyperparameter that is frequently fixed prior to model-fitting. The predictive distribution is typically assumed to be normal \citep{gal2016dropout}, hence in practice 2 standard deviations are applied when constructing prediction intervals. However, this approach often fails to capture true model uncertainty. Past research \citep{liu2021peril, laves2019well} demonstrates that MC Dropout is prone to miscalibration, meaning the predicted probabilities do not reliably reflect the actual error rates.

\section{Cubing strategy for stable subregion identification}
\label{sec:Cubing}
In MC dropout, the quality of the resulting predictive intervals depends critically on several hyperparameters, particularly the dropout rate ($p$), weight decay ($\lambda$), and the multiplier used to construct prediction intervals ($k$). In practice, these quantities are often fixed heuristically or tuned independently, which can lead to poorly calibrated uncertainty estimates \citep{wu2021quantifying, zhu2017deep, mehrtash2020confidence}. Moreover, even small changes in these parameters can substantially alter the variance of the predictive distribution \citep{zeevi2025rate, de2025deep}, producing intervals that are either overly narrow or excessively wide. Exhaustively searching the hyperparameter space can be computationally prohibitive for large spatial datasets since each candidate configuration requires training a separate neural network. To address this challenge, we propose a cubing strategy that identifies stable regions of the hyperparameter space where MC dropout yields well-calibrated predictive intervals. Rather than targeting a single optimal configuration, the proposed approach searches for regions in which predictive performance consistently matches or exceeds that of a baseline model. 
\subsection{Hyperparameter Search Formulation}
Let $\mathcal{H} = \Lambda \times P \times K$ denote the hyperparameter space, where $\Lambda$ represents the weight decay parameter, $P$ denotes the dropout rate, and $K$ denotes the standard deviation multiplier used in constructing prediction intervals. Each hyperparameter configuration can be written as
$\mathbf{h} = (\lambda, p, k) \in \mathcal{H}$. For a given configuration $\mathbf{h}$, a neural network is trained using the corresponding hyperparameters, and predictive performance (accuracy and uncertainty quantification) is evaluated on a holdout validation set using scoring rules, such as the mean interval score (MIS) and the continuous ranked probability score (CRPS). Details regarding these scores can be found in the Supplement Section~\ref{scoringrulesdef}. 

Let $S(\mathbf{h})$ and $S_{\text{base}}$ denote the score corresponding to configuration $\mathbf{h}$ and a baseline model, such as the Bayesian SLMM with basis representation in \eqref{eq:basis_slmmm}. We use a Bayesian SLMM as the baseline due to its well-established inferential properties in spatial regression. Note that the baseline serves as a calibration anchor rather than an optimal benchmark. We define the stable hyperparameter set $\mathcal{R}$ as the set of configurations yielding predictive intervals whose performance is at least as good as the baseline model:
\[
\mathcal{R}
:=
\left\{ \mathbf{h} \in \mathcal{H} : S(\mathbf{h}) \le S_{\text{base}} \right\}.
\]

\subsection{Cubing Strategy}
Direct estimation of $\mathcal{R}$ is challenging since the hyperparameter space $\mathcal{H}$ may be high-dimensional and the subset satisfying $S(\mathbf{h}) \le S_{\text{base}}$ may exhibit complex geometry. In practice, $\mathcal{R}$ may be fragmented, composed of several irregular and nonconvex subregions. 

To address this problem, we approximate $\mathcal{R}$ using a collection of axis-aligned hyperrectangles, referred to as cubes for simplicity. An axis-aligned cube restricts each hyperparameter component ($\lambda, p, k$) independently within an interval and has boundaries that follow the coordinate axes of the hyperparameter space. The cube takes the form
\[
\mathcal{C} = \{ \mathbf{h} \in \mathcal{H} : a_j \le h_j \le b_j,\; j=1,2,3\},
\]
whose faces are parallel to the coordinate axes of the hyperparameter space. In our setting, the hyperparameters are three-dimensional, so the cubes correspond to rectangular subregions in $\mathbb{R}^3$. We approximate $\mathcal{R} \approx \bigcup_{c=1}^{C} \mathcal{C}_c$
where each $\mathcal{C}_c$ represents a candidate cube. 

The objective is to identify a collection of cubes $\mathcal{C}_c$ containing hyperparameter configurations whose predictive performance matches or improves upon that of the baseline model (e.g., results from the Bayesian SLMM). The baseline serves as a calibration anchor used to identify regions where predictive uncertainty is reliably calibrated, as opposed to an optimal benchmark. Let $\mathcal{H}(C)$ denote the set of hyperparameter configurations contained within cube $C$. The average score of the cube is defined as
\[
\overline{S}(C)
=
\frac{1}{|\mathcal{H}(C)|}
\sum_{h \in \mathcal{H}(C)} S(h),
\]
and the baseline beat rate is defined as
\[
w(C)
=
\frac{1}{|\mathcal{H}(C)|}
\sum_{h \in \mathcal{H}(C)} \mathbb{I}\{S(h) < S_{\text{base}}\}.
\]

The quantity $\overline{S}(C)$ measures the average predictive performance of hyperparameters within the cube and $w(C)$ represents the proportion of configurations that outperform the baseline model. We combine these measures into an overall cube score
\[
O(C) = w(C) + (1 - \overline{S}(C)),
\]
which prioritizes cubes containing hyperparameter configurations with both strong predictive performance. Other scoring functions could also be used, but we adopt this form for interpretability.

The cubing procedure recursively partitions the hyperparameter space into smaller cubes and evaluates each region using the score $O(C)$. Cubes that consistently outperform the baseline are retained, while subregions with subpar predictive performance are discarded. This process yields a compact representation of stable hyperparameter subregions that can help guide future hyperparameter tuning.

\paragraph{Cubing Algorithm}
The cubing strategy is implemented using the recursive search procedure described in Algorithm~\ref{alg_cube_search}. The algorithm requires several user-specified inputs including: 
(i) a baseline score $B$ (either $\mathrm{MIS}_{\text{base}}$ or $\mathrm{CRPS}_{\text{base}}$); 
(ii) an initial cube size $s_0$ defining the width, length, and height of the search region; 
(iii) a scaling rule, denoted \textsc{NextSize}, which determines the size of successive cubes by halving each dimension of the current cube; 
(iv) a maximum number of refinement rounds $T_{\max}$; 
(v) an improvement window $t$ specifying the number of iterations used to assess score improvement; and 
(vi) a stopping threshold $\epsilon$, defined as the percentage change in the score relative to the previous iteration used to assess convergence.

Before running the algorithm, a neural network is trained for each $(p,\lambda) \in \mathcal{H}$, while prediction intervals are evaluated across candidate values of $k$. For each trained network and every $k \in \mathcal{H}$, the score $S(p,\lambda,k)$ is evaluated on the holdout set. The evaluated hyperparameter combinations define a ``hyperparameter box''
\[
\mathcal{B} = 
[\min \lambda,\max \lambda] \times
[\min p,\max p] \times
[\min k,\max k],
\]
constructed as the smallest axis-aligned region that bounds all explored triples $(p,\lambda,k)$. During the search, candidate cubes are evaluated relative to their parent cube using a score $O(C)$.

\begin{algorithm}[H]
\caption{Cube Search for MC Dropout Hyperparameters}
\label{alg_cube_search}
\begin{algorithmic}[1]

\State Initialize queue $\mathcal{Q}$ with the starting cube $C_0$ covering the hyperparameter box $\mathcal{B}$.
\State Set initial cube size $s \gets s_0$.

\While{$\mathcal{Q}$ is not empty and $t_{\text{round}} \le T_{\max}$}

\State Select cube $C$ from $\mathcal{Q}$.
\State Let $\mathcal{H}(C)$ be the set of evaluated hyperparameter combinations in $C$.
\State Compute mean score
\[
\overline{S}(C) =
\frac{1}{|\mathcal{H}(C)|}
\sum_{h\in\mathcal{H}(C)} S(h).
\]

\State Compute baseline beat rate
\[
w(C) =
\frac{1}{|\mathcal{H}(C)|}
\sum_{h\in\mathcal{H}(C)} \mathbb{I}\{S(h) < B\}.
\]

\State Compute cube score
\[
O(C) = w(C) + (1 - \overline{S}(C)).
\]

\State \textbf{Early stopping:}
If the relative improvement between $O(C)$ and its parent cube remains below threshold $\epsilon$ for $t$ successive ancestors, stop splitting.

\State Otherwise, split cube $C$ into smaller subcubes using
\[
s' \gets \textsc{NextSize}(C,s).
\]

\State Add resulting subcubes to queue $\mathcal{Q}$.

\EndWhile

\State \textbf{return} the top $k$ cubes ranked by $O(C)$.

\end{algorithmic}
\end{algorithm}

\begin{figure}[H]
     \centering
     \includegraphics[width=1\linewidth]{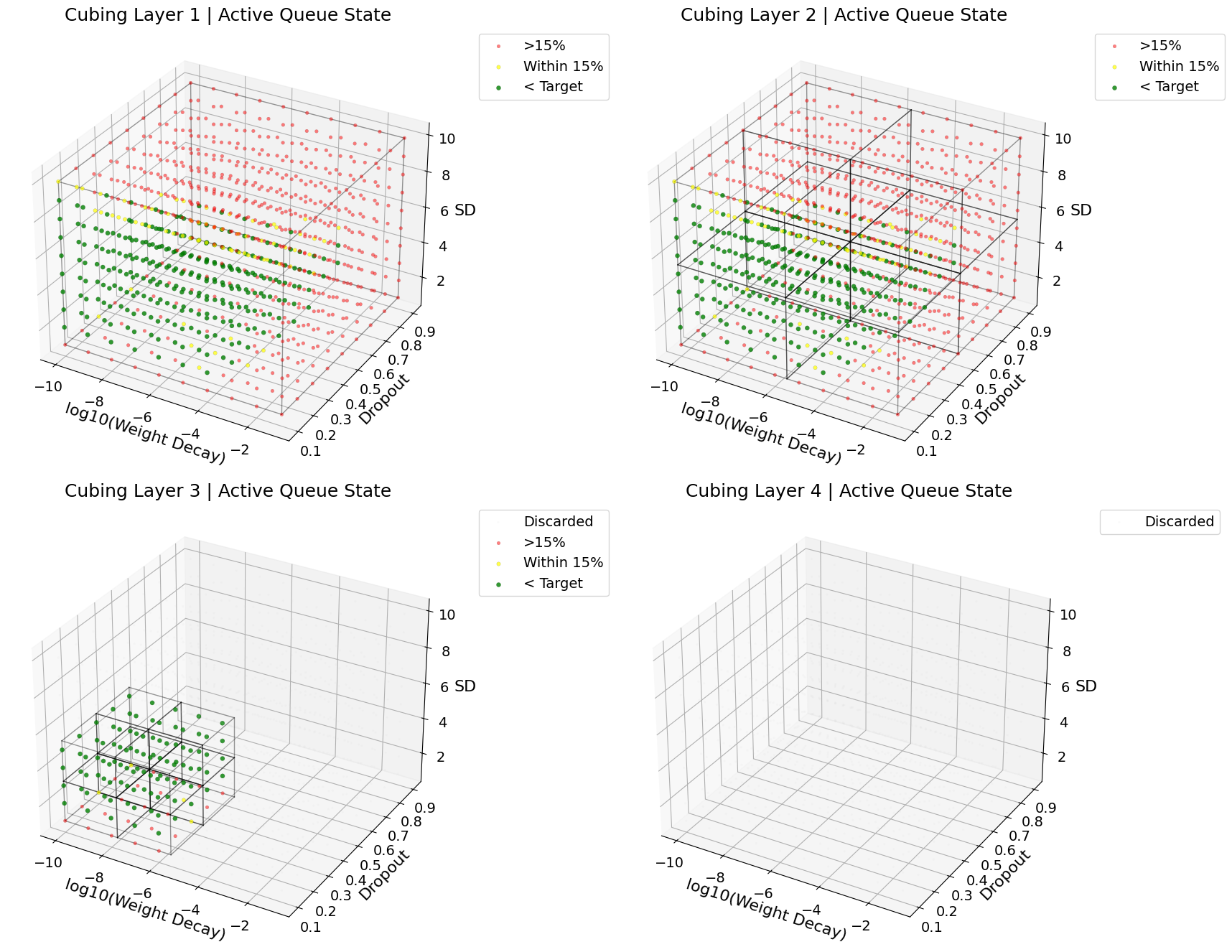}
     \caption{Example execution of the cubing algorithm ($\epsilon=0.3$, max improvement window=1).}
     \label{fig:placeholder}
 \end{figure}

\paragraph{Demonstration}
We provide the following proof-of-concept of the proposed cubing algorithm, with results displayed in Figure ~\ref{fig:placeholder}. At the first step of the algorithm (Layer 1), the entire hyperparameter space is treated as a single cube. Each point represents a unique combination of weight decay, sd multiplier, and dropout. The points are colored based on their performance relative to the baseline model: green indicates MIS lower than the baseline (here, the linear SLMM), yellow indicates values within 15\% of the baseline score, and red indicates values exceeding the baseline by more than 15\%. Here, a cluster of points emerges, which satisfy the green criteria. In the next step (Layer 2), the cubes have been generated by splitting across each of the dimensions. Note that some cubes contain mostly green points so they meet the $\epsilon$ criteria of improvement from the parent cube (in this case the entire hyperparameter space) and thus have their window still at 0. Some other cubes have almost all red points and have not actually improved from their parent cube and are currently at the max improvement window. In the third step (Layer 3), after each of the cubes have been processed at the next cube layer the only remaining sub cubes are those whose parent cubes had actually improved, in the case of cubes that dropped this is because splitting had not improved upon the average score in the parent cube. In the fourth step (Layer 4) splitting the remaining sub cubes from cubing layer 3 does not improve upon the score and thus those are now dropped into the finished cubes set. From here each of the processed cubes are logged in order to compare with other runs of the algorithm to identify the common top cubes across runs.

\section{Simulation Study}
\label{simstudy}
We evaluate the predictive accuracy and uncertainty quantification of the proposed cubing-based NB-SLMM under varying spatial dependence regimes defined by the smoothness $\nu$ and range $\rho$ parameters of the Mat\'ern covariance function in \eqref{eq:Matern}. Two cubing strategies are considered: one that identifies stable subregions using the mean interval score (MIS), and another based on the continuous ranked probability score (CRPS). Performance is compared with a fully Bayesian hierarchical SLMM using the basis representation described in Section~\ref{subsec:basis_SLMM}.

\subsection{Study Design}

We randomly generate spatial locations $s_i \in \mathcal{D}=[0,1]^2 \subset \mathbb{R}^2$ for $i=1,\ldots,N$, where $\mathcal{D}$ denotes the spatial domain. Each dataset contains $N=25{,}000$ locations, which are partitioned into $N_{\text{train}}=20{,}000$ training and $N_{\text{test}}=5{,}000$ testing observations. The response vector $\mathbf{Z}=(Z(s_1),\ldots,Z(s_N))^\top$ is generated from the spatial linear mixed model in \eqref{eq:SLMM}. Covariates $\mathbf{X}=[X_1,X_2]$ are generated independently from $\text{Unif}(-0.5,0.5)$ with regression coefficients $\boldsymbol{\beta}=(1,1)^\top$. Spatial random effects are simulated from a zero-mean Gaussian process with a Mat\'ern covariance function. We consider four spatial dependence settings defined by smoothness parameters $\nu\in\{0.5,1.5\}$ and effective ranges $\{0.3,0.6\}$. The effective range is the distance at which spatial correlation decays to $0.05$, and we select the corresponding range parameters to achieve these values. For each setting, we generate 25 replicate datasets, resulting in a total of 100 simulated datasets. 

To construct the basis functions matrix $\boldsymbol{\Phi}$, we use the leading eigenvectors of a Mat\'ern covariance matrix evaluated at the observed locations. For setting $i$, we use the covariance parameters $\nu_i\in\{0.5,1.5\}$ and effective ranges $\rho_{eff,i}\in\{0.3,0.6\}$ that were used to generate the spatial random effects. Then, we construct an $N\times N$ Mat\'ern covariance matrix $\tilde{\boldsymbol{\Sigma}}_{\tilde{\boldsymbol{\theta}}_i}$ with covariance parameter vector
$\tilde{\boldsymbol{\theta}}_i=(\sigma^2=1, \rho_{eff} = \rho_{eff,i}, \nu= \nu_i)^\top$.  To illustrate, for setting 1, we generate the data using $\nu_1=0.5$ and $\rho_{eff,i}=0.3$, and these exact parameters are used to generate the associated covariance matrix $\tilde{\boldsymbol{\Sigma}}_{\tilde{\boldsymbol{\theta}}_i}$. From the eigendecomposition $\tilde{\boldsymbol{\Sigma}}_{\boldsymbol{\theta}_i} \approx \mathbf{U}_m \boldsymbol{\Lambda}_m \mathbf{U}_m^\top$, we obtain the basis matrix as $\boldsymbol{\Phi}=\mathbf{U}_m$. 
This construction enables us to study model performance across varying spatial dependence structures using basis representations of dimension $m\in\{25,135,250\}$.

For the baseline model, we use the Bayesian hierarchical model described in Section~\ref{subsec:basis_SLMM}. Posterior inference and predictions are based on samples drawn using the Metropolis–Hastings algorithm. The sampler is run for 50,000 iterations following a 5,000-iteration burn-in period. We employ a normal prior for $\tilde{\boldsymbol{\beta}}\sim \mathcal{N}(\mathbf{0},100\mathbf{I})$ alongside an inverse-gamma prior for $\tilde{\tau}^2\sim \mathcal{IG}(2,\sigma^2_{train})$. We draw samples from the posterior predictive distributions to evaluate the Continuous Ranked Probability Score (CRPS) and Mean Interval Score (MIS) at the test locations. All computations were performed on the HOPPER high-performance computing system at George Mason University using Intel Xeon Gold 6240R processors.

\paragraph{Cubing Algorithm Design}
MC Dropout produced prediction intervals with the predictive standard deviation multipliers ranging from 1 to 10. Neural networks were trained across a grid, with weight decay values ranging from $10^{-10}$ to $10^{-1}$ and dropout rates ranging from 0.1 to 0.9. The cubing algorithm was applied in each of the 25 replicates with the hyperparameter subregion being formed from the top 5 performing cubes across all replicates. Each hyperparameter subregion consists of combining the top 5 cubes that had the highest amount of top 10 performance rankings in each of the 25 replicates. Each run of the cubing algorithm used a max round of $10,000, \epsilon = 0.15$, and $t=3$. We report the RMSE, MIS, prediction interval width, as well as $N_{\text{top}}$. In order to show consistency of the cubes within the estimated subregion, the quantity $N_{\text{top}}$ represents the average number of times a cube within the selected subregion appears among the top 10 performing hyperparameter configurations across the 25 replicates.






\subsection{Simulation Study Results}\label{sec:sim_results}

Overall, the learned-aleatoric (LA) MC dropout model provides the best balance of predictive accuracy, interval calibration, and hyperparameter stability across the simulation settings. We therefore focus on the Bayesian regression baseline and the best-performing LA MC dropout model, as these represent the primary comparisons of interest. Across methods, predictive performance generally improves as the number of basis functions $m$ increases, with the largest gains occurring between $m=25$ and $m=135$. Performance also improves in smoother spatial settings with larger effective ranges (i.e., higher $\nu$ and $\rho$), suggesting that stronger spatial structure is easier to capture across modeling approaches. Additional results, including alternative MC dropout formulations, hyperparameter subregion summaries, and CRPS-based evaluations, are provided in Supplement~\ref{app:simulation}.


\paragraph{Spatial Bayesian Regression Results}
\label{slr_res_mis}

The Bayesian regression baseline results in Table~\ref{bayes_base} show improvement in predictive performance as the number of basis functions $m$ increases. Across all four settings, increasing $m$ leads to reductions in MIS, RMSE, and interval width, indicating both improved accuracy and lower prediction errors. The most substantial gains occur when moving from $m=25$ to $m=135$, with more modest improvements from $m=135$ to $m=250$, suggesting diminishing returns beyond a moderate number of basis functions. Coverage remains close to the nominal 95\% level across all settings and values of $m$, indicating that the Bayesian regression model is well-calibrated despite differences in spatial structure.

\begin{table}[H]
\centering
\begin{tabular}{rrrrrr}
\toprule
   Set &   m &   MIS &   RMSE &   Width &    Cvg \\
\midrule
     1 &  25 & 3.642 &  0.777 &   3.046 & 0.949 \\
     1 & 135 & 2.561 &  0.546 &   2.138 & 0.949 \\
     1 & 250 & 2.282 &  0.488 &   1.918 & 0.950 \\
     \midrule
     2 &  25 & 3.267 &  0.697 &   2.731 & 0.949 \\
     2 & 135 & 1.551 &  0.331 &   1.292 & 0.948 \\
     2 & 250 & 1.33  &  0.284 &   1.11  & 0.948 \\
     \midrule
     3 &  25 & 2.861 &  0.615 &   2.418 & 0.951 \\
     3 & 135 & 1.984 &  0.423 &   1.659 & 0.950 \\
     3 & 250 & 1.787 &  0.381 &   1.492 & 0.949 \\
     \midrule
     4 &  25 & 1.949 &  0.414 &   1.628 & 0.950 \\
     4 & 135 & 1.086 &  0.232 &   0.911 & 0.950 \\
     4 & 250 & 1.092 &  0.234 &   0.915 & 0.949 \\
\bottomrule
\end{tabular}
\caption{\textbf{Bayesian regression baseline across spatial settings.}
Results are shown for four data-generating settings defined by smoothness $\nu$ and effective range $\rho$: (1) $\nu=0.5$, $\rho=0.3$; (2) $\nu=1.5$, $\rho=0.3$; (3) $\nu=0.5$, $\rho=0.6$; (4) $\nu=1.5$, $\rho=0.6$. The parameter $m$ denotes the number of spatial basis functions. Metrics include mean interval score (MIS), root mean squared error (RMSE), interval width (Width), and coverage (Cvg, \%)}
\label{bayes_base}
\end{table}

\paragraph{MC Dropout Results}

In contrast, MC Dropout with EU produces narrower intervals but suffers from undercoverage, reflecting insufficient uncertainty quantification. The FA method with MC Dropout improves calibration and stabilizes hyperparameter selection, though coverage remains somewhat variable relative to the Bayesian baseline. 

The LA MC Dropout approach, which explicitly models input-dependent aleatoric variance, achieves the most favorable trade-off between interval width and coverage. Across nearly all settings and values of $m$, it produces intervals that are both sharper than the Bayesian baseline and well-calibrated, often achieving or slightly exceeding nominal coverage. As a result, it consistently attains the lowest MIS, RMSE, width values, indicating superior interval quality and prediction performance.

\begin{table}[ht]
\centering
\begin{tabular}{rrrrrr}
\toprule
   Set &   m &   MIS &   RMSE &   Width &   Cvg \\
\midrule
     1 &  25 &  2.324 &  0.478 &   2.165 & 0.978 \\
     1 & 135 &  1.905 &  0.395 &   1.624 & 0.957 \\
     1 & 250 &  1.73  &  0.364 &   1.446 & 0.95  \\
     \midrule
     2 &  25 &  1.808 &  0.352 &   1.749 & 0.987 \\
     2 & 135 &  1.215 &  0.255 &   1.083 & 0.962 \\
     2 & 250 &  1.454 &  0.303 &   1.298 & 0.964 \\
     \midrule
     3 &  25 &  1.826 &  0.37  &   1.734 & 0.98  \\
     3 & 135 &  1.661 &  0.35  &   1.424 & 0.955 \\
     3 & 250 &  1.496 &  0.319 &   1.337 & 0.962 \\
     \midrule
     4 &  25 &  1.426 &  0.285 &   1.34  & 0.977 \\
     4 & 135 &  0.936 &  0.193 &   0.853 & 0.971 \\
     4 & 250 &  1.095 &  0.229 &   0.982 & 0.965 \\
\bottomrule
\end{tabular}
\caption{\textbf{Performance of MC Dropout with LA across spatial settings.}
Results are reported across four data-generating settings and basis dimensions $m \in \{25, 135, 250\}$, where $m$ denotes the number of leading eigenvectors used to represent spatial basis functions. The four settings correspond to combinations of smoothness $\nu$ and effective range $\rho$: (1) $\nu=0.5$, $\rho=0.3$; (2) $\nu=1.5$, $\rho=0.3$; (3) $\nu=0.5$, $\rho=0.6$; (4) $\nu=1.5$, $\rho=0.6$. For each setting and $m$, performance is evaluated using mean modified interval score (MIS), root mean squared error (RMSE), average prediction interval width (Width), and empirical coverage (Cvg). Results are averaged over 25 replicates.}
\label{mc_drop_pred_res}
\end{table}

\begin{table}[ht]
\centering
\begin{tabular}{rrlllr}
\toprule
   Set &   m & WDR                & DR           & SDR          &   Ntop \\
\midrule
     1 &  25 & (1.0e-10, 3.2e-06) & (0.10, 0.15) & (1.56, 2.12) &   21.6 \\
     1 & 135 & (3.7e-10, 1.2e-05) & (0.10, 0.25) & (1.56, 2.12) &    9.6 \\
     1 & 250 & (6.5e-08, 1.5e-04) & (0.10, 0.15) & (1.56, 2.12) &   14.6 \\
     \midrule
     2 &  25 & (4.9e-09, 1.2e-05) & (0.10, 0.15) & (1.56, 2.12) &   22.6 \\
     2 & 135 & (8.7e-07, 1.5e-04) & (0.10, 0.15) & (1.56, 2.12) &   20   \\
     2 & 250 & (3.2e-06, 2.1e-03) & (0.10, 0.15) & (1.56, 2.12) &   18   \\
     \midrule
     3 &  25 & (3.7e-10, 1.2e-05) & (0.10, 0.15) & (1.56, 2.12) &   22.2 \\
     3 & 135 & (4.9e-09, 1.5e-04) & (0.10, 0.15) & (1.56, 2.12) &   17.4 \\
     3 & 250 & (4.9e-09, 2.1e-03) & (0.10, 0.15) & (1.56, 2.12) &   17.2 \\
     \midrule
     4 &  25 & (4.9e-09, 1.2e-05) & (0.10, 0.15) & (1.56, 2.12) &   20.2 \\
     4 & 135 & (3.7e-10, 1.5e-04) & (0.10, 0.15) & (1.56, 2.12) &   20.8 \\
     4 & 250 & (6.5e-08, 1.5e-04) & (0.10, 0.15) & (1.56, 2.12) &   22.2 \\
\bottomrule
\end{tabular}
\caption{\textbf{Top-performing hyperparameter subregions for MC Dropout with LA.}
For each spatial setting and basis dimension $m \in \{25, 135, 250\}$, the table reports the  hyperparameter subregion obtained by combining the top 5 performing cubes across 25 replicates using the cubing algorithm. The four settings correspond to combinations of smoothness $\nu$ and effective range $\rho$: (1) $\nu=0.5$, $\rho=0.3$; (2) $\nu=1.5$, $\rho=0.3$; (3) $\nu=0.5$, $\rho=0.6$; (4) $\nu=1.5$, $\rho=0.6$. WDR denotes the weight decay range, DR the dropout rate range, and SDR the predictive standard deviation multiplier range. The quantity $N_{\text{top}}$ represents the average number of times a cube within the selected subregion appears among the top 10 performing hyperparameter configurations across the 25 replicates. Results are based on MIS with $\gamma=1$, assigning equal importance to interval width and coverage.}
\label{mc_drop_pred}
\end{table}

This approach aligns with the standard approach of using approximately two standard deviations for interval construction with MC Dropout \citep{gal2016dropout, wang2024uncertainty, kummaraka2025monte}. This is reflected in the narrow SDR ranges identified through cubing, which also uses two standard deviations; whereas the EU and FA approaches which required much higher SDR ranges to contend with the baseline approach.

Dropout ranges are stable across subregions, with nearly all subregions favoring values between $0.10$ and $0.15$, suggesting that only modest stochastic regularization is needed when both epistemic and aleatoric components are considered. Weight decay ranges exhibit some variation, particularly as $m$ increases, but remain within relatively narrow bands. CRPS-based results outperformed the baseline spatial Bayesian Regression model, with the LA modeling approach performing the best. A key distinction is that the identified hyperparameter subregions tended to favor higher dropout ranges and lower weight decay ranges, likely due to the absence of tuning for standard deviation to control for capturing coverage. See Subsection~\ref{alg_res_crps} of the Supplement for additional details. 


Our simulation results reveal that the choice of uncertainty modeling strategy may be as important as the hyperparameter search itself. In particular, the LA MC Dropout model, which explicitly learns input-dependent aleatoric variance, consistently provides the best performance, as evidenced by the MIS and CRPS scores, while also yielding highly stable hyperparameter regions. In contrast, the EU MC Dropout model often required large SDR values to compensate for missing aleatoric uncertainty. The FA approach, while generally providing better results than baseline, still depends on the interaction between dropout-induced epistemic variability and the added fixed variance term. These results suggest that explicitly learning aleatoric uncertainty provides a structural improvement for predictive uncertainty quantification in spatial deep learning.

\section{Application: Remotely Sensed Land Surface Temperature}
\label{sec:application}
NASA’s Terra satellite, part of the Earth Observing System Morning Constellation (EOS-AM), was launched on December 18, 1999, and has provided global observations of the Earth’s atmosphere, oceans, and land surface since February 2000. In this study, we analyze land surface temperature (LST) derived from the Moderate Resolution Imaging Spectroradiometer (MODIS) instrument aboard the Terra satellite. The LST data are obtained from the \texttt{MOD11\_L2} product collected on July 1, 2025, at approximately 1:00 PM local time. The study region spans longitudes $16.5^{\circ}$–$24.0^{\circ}$ E and latitudes $36.0^{\circ}$–$41.0^{\circ}$ N, comprising $215{,}941$ grid cells at a spatial resolution of 1 km. Since LST values (in $^\circ$C) are strictly positive, we apply a logarithmic transformation prior to model fitting. For computational feasibility, we randomly selected $n = 30{,}000$ locations for model fitting and reserved an additional $20\%$ of the locations as a holdout set for validation. The neural network architecture and training procedures follow those used in the simulation study (Section~\ref{simstudy}). For the spatial basis functions, we employ the leading $m=\{20,135,250\}$ eigenvectors of two pre-specified Mat\'ern covariance matrices. The first is a Mat\'ern covariance matrix with $\nu_1=0.5$ and the range parameter $\rho_1$ was inferred using the \texttt{fields} package \citep{furrer2009package} on a smaller subset of the dataset. We do the same for the second case where $\nu_2=1.5$ to obtain $\rho_2$. We provide more details in the Supplement~\ref{supp:basis} about basis generation. Additional information about basis generation is provided in Section~\ref{supp:basis} of the Supplement.

In this study, we elected to transfer the stable subregions learned from the simulation study in Section~\ref{simstudy} based on comparable spatial dependence settings. To that end, we link the tuples $(\nu_1,\rho_1)$ and $(\nu_2,\rho_2)$ to the appropriate simulation setting and hyperparameter subregion combination from Section~\ref{sec:sim_results}. Specifically, $(\nu_1,\rho_1)$ corresponds to setting 3 (rougher process with longer-range dependence) and $(\nu_2,\rho_2)$ to setting 2 (moderate smoothness with short-range dependence). 

We present results for the case $(\nu_1,\rho_1)$ as it consistently achieved better scores under the full profile likelihood and REML criteria. Results for the alternative specification $(\nu_2,\rho_2)$ were deferred to Supplement~\ref{alternate_nu}.


\begin{table}[H]
\centering
\begin{tabular}{lrrrrrr}
\toprule
 Model    &   m &   MIS &   CRPS &   RMSE &   Width &   Cvg \\
\midrule
  Base & 20   & 0.36  &  0.037 &  0.026 &   0.347 & 0.997 \\
 Base & 135   & 0.347 &  0.028 &  0.023 &   0.337 & 0.998   \\
 Base & 250   & 0.345 &  0.026 &  0.023 &   0.335 & 0.998   \\
 \midrule
 EU &  20 & 0.214 &  0.014 &  0.028 &   0.181 & 0.978 \\
 EU & 135 & 0.158 &  0.011 &  0.023 &   0.128 & 0.979 \\
 EU & 250 & 0.435 &  0.015 &  0.023 &   0.412 & 0.985 \\
 \midrule
 FA    &  20 & 0.472 &  0.014 &  0.028 &   0.462 & 0.997 \\
 FA    & 135 & 0.458 &  0.01  &  0.022 &   0.449 & 0.998 \\
 FA    & 250 & 0.458 &  0.01  &  0.022 &   0.449 & 0.998 \\
 \midrule
 LA &  20 & 0.155 &  0.014 &  0.029 &   0.137 & 0.991 \\
 LA & 135 & 0.104 &  0.01  &  0.022 &   0.088 & 0.986 \\
 LA & 250 & 0.112 &  0.009 &  0.022 &   0.097 & 0.988 \\
\bottomrule
\end{tabular}
\caption{\textbf{Performance for the LST application under Setting 3 ($\nu=0.5$, long-range dependence).}
Results are reported for baseline and the three MC Dropout variants: EU, FA, and LA. Performance is evaluated across basis dimensions $m \in \{20, 135, 250\}$, where $m$ denotes the number of spatial basis functions. Metrics include mean interval score (MIS), continuous ranked probability score (CRPS), root mean squared error (RMSE), average prediction interval width (Width), and empirical coverage (Cvg). Results are computed on the held-out validation set.}
\label{applied_results_0_5}
\end{table}

Results in Table~\ref{applied_results_0_5} show consistent patterns across modeling approaches under Setting 3. In both the baseline model and the three MC Dropout approaches, the baseline model exhibits extreme over-coverage (approximately $99.8\%$), indicating substantial over-dispersion in the predictive distribution and resulting in unnecessarily wide intervals and elevated MIS values.

Across all methods, increasing the number of basis functions generally improves RMSE and CRPS, reflecting better representation of the spatial field. With more basis functions incorporated, particularly from $135$ to $250$, the EU MC Dropout model substantially increases in width and MIS but not CRPS. The FA approach produces wider intervals and higher MIS values relative to the MC Dropout EU. This reflects persistent overestimation of predictive uncertainty, although with a maximum MIS $\sim 0.94$ and minimum MIS $\sim 0.16$, the identified hyperparameter subregions still contain results superior to the baseline Bayesian regression model. While both MC Dropout with EU and FA provide improvements over Bayesian regression in terms of predictive accuracy and interval quality, the LA approach offers the best balance across uncertainty measures (MIS, CRPS, RMSE, prediction interval width, and coverage) for all basis dimensions of $m$.

\section{Discussion}\label{sec:discussion}
This study develops a framework for uncertainty quantification in spatial deep learning models by introducing a computationally-efficient cubing-based strategy to identify stable hyperparameter regions for MC dropout. As opposed to ad hoc or exhaustive tuning, the cubing approach systematically partitions the hyperparameter space and identifies compact regions that yield well-calibrated predictive intervals relative to a statistical baseline. Across both simulation studies and a large remotely sensed land surface temperature dataset, the results show that the resulting subregions preserve strong-performing configurations while substantially shrinking the search space. In many cases, the identified regions even outperform the Bayesian regression benchmarks obtained via MCMC. Our simulation results show that the choice of uncertainty modeling strategy (EU, FA, and LA) may be as important as hyperparameter tuning, with explicit modeling of aleatoric variance (LA) improving both accuracy and interval estimation. Additionally, our results still provide subregions with the EU and FA cases that can generally outperform the baseline methods. These results provide practical guidance for tuning MC Dropout so that NB-SLMM models yield reliable uncertainty quantification. 



Several limitations should be noted. For MC Dropout, the effective number of parameters increases considerably as the number of basis functions $m$ grows, relative to the sample size $n$. For example, when $m=250$, the resulting network contains 33,261 trainable parameters for only 16,000 training points. Although neural networks can sometimes perform well even with a relatively small data-to-parameter ratio \citep{allen2019learning}, this imbalance may still contribute to the instability observed at larger basis dimensions. Next, further work on both simulated and real datasets with much larger $N$ would be useful for clarifying how these methods scale and whether the advantages of the reduced cubing regions become even more pronounced in large-sample regimes. We also note that the coarse hyperparameter grid and axis-aligned cubing restrict the ability to capture irregularly-shaped regions that are high-performing. Finer grids could enable more flexible methods such as clustering-based approaches. In this study, our results are based on a single class of spatial basis functions, yet further exploration is required to assess whether the identified subregions are comparable across different basis functions. The real-world application suggests that stable hyperparameter regions may transfer across similar settings. A broader empirical study is needed to determine when the cubing procedure must be rerun and when previously identified stable regions can be reused, which could inform practical guidelines and heuristics.

\section*{Acknowledgements}\label{Sec:Acknowledgements}This project was supported by computing resources provided by the Office of Research Computing at George Mason University (\url{https://orc.gmu.edu}) and funded in part by grants from the National Science Foundation (Awards Number 1625039 and 2018631). 

\newpage








\bibliographystyle{elsarticle-harv}
\bibliography{bibfile}
\addcontentsline{toc}{section}{References} 

\clearpage

\section*{Supplementary Material for ``A Cubing Strategy for Identifying Stable Hyperparameter Regions for Uncertainty Quantification in Spatial Deep Learning''}
\setcounter{page}{1}
\setcounter{section}{0}
\renewcommand{\thesection}{\arabic{section}}

\section{Implementation Details}
\label{supp:impl}
This section describes the neural network architecture used in the NB-SLMM and the associated tuning parameters.
\subsection{Neural Network Architecture}
\label{NN_arc}
\begin{figure}[H]
    \centering
    \includegraphics[width=0.44\linewidth]{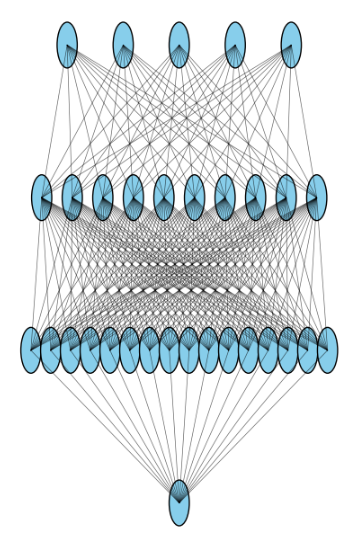}
    \caption{Illustration of the neural network architecture for an example with five input features.}
    \label{fig:nn_ex_5_input}
\end{figure}

The neural network architecture used in this study consists of an input layer, two hidden layers, and an output layer, as illustrated in Figure~\ref{fig:nn_ex_5_input}.

\noindent\textbf{Input layer.} The input layer contains $p$ neurons, where $p$ denotes the number of covariates together with the eigenvector-based spatial basis functions used to approximate the unobserved spatial process. Each input vector corresponds to
\[
\mathbf{x}_\phi(\mathbf{s}) =
\big(\mathbf{x}(\mathbf{s})^\top,\boldsymbol{\phi}(\mathbf{s})^\top\big)^\top .
\]

\noindent\textbf{Hidden layer 1.} The first hidden layer contains
\[
h_1 = \min(2p,\,100)
\]
neurons. This layer expands the feature representation of the inputs and captures nonlinear relationships between covariates and spatial basis features while preventing excessive width when $p$ is large.

\noindent\textbf{Hidden layer 2.} The second hidden layer contains
\[
h_2 = \max\big(\lfloor 0.8 h_1 \rfloor,\,16\big)
\]
neurons. This layer produces a compressed representation of the features learned in the first hidden layer, enabling hierarchical feature extraction.

\noindent\textbf{Output layer.} The output layer consists of a single neuron with identity activation, producing the predicted response
\[
\hat{z}(\mathbf{s}) = f(\mathbf{x}_\phi(\mathbf{s});\boldsymbol{\psi}).
\]
This layer maps the learned features from the final hidden layer to the scalar regression output.

\subsection{Regularization and Tuning Parameters}

Key hyperparameters influencing uncertainty quantification in MC dropout include: (i) the dropout rate $p$; (ii) weight decay $\lambda$; and (iii) and the assumed predictive variance $\sigma^2_{tot}$. 
Both weight decay and dropout address overfitting but act differently. Weight decay imposing shrinkage \citep{bishop2006pattern,krogh1991simple}, whereas dropout injects multiplicative noise into activations to approximate model averaging \citep{srivastava2014dropout}. 

The dropout rate influences the variability of the sampled subnetworks; hence, it is an important hyperparameter affecting epistemic uncertainty. Previous work has shown that fixing this parameter heuristically can lead to unstable uncertainty estimates \citep{wu2021quantifying,zhu2017deep}. 

Weight decay interacts closely with the dropout rate through the relationship $\lambda = \frac{p\,\ell^{2}}{2N\tau}$ in the MC dropout framework \citep{gal2016dropout}. Consequently, varying the dropout rate while holding $\lambda$ fixed implicitly alters the prior precision of the model, affecting how the network separates signal from noise. Recent studies indicate that common implementations of MC dropout may suffer from over-regularization of weights $\mathbf{W}$, leading to overly similar sampled subnetworks and overconfident predictions \citep{liu2021peril,de2025deep}, even when the predictions are inaccurate. 

Standard deviations is another hyperparameter that is frequently fixed. The predictive distribution is typically assumed to be normal \citep{gal2016dropout}, hence in practice 2 standard deviations are applied when constructing prediction intervals. However, this approach often fails to capture true model uncertainty; research \citep{liu2021peril, laves2019well} demonstrates that MC Dropout is prone to miscalibration, meaning the predicted probabilities do not reliably reflect the actual error rates.

\section{Priors}
\label{supp:prior}
\paragraph{Flat prior for $\boldsymbol{\beta}$.}
We first assume a noninformative (flat) prior for $\boldsymbol{\beta}$, i.e.,
\[
p(\boldsymbol{\beta}) \propto 1.
\]
Conditional on $\sigma^2$, $\mathbf{X}$, and $\mathbf{Y}$, the full conditional distribution is
\[
\boldsymbol{\beta} \mid \sigma^2, \mathbf{X}, \mathbf{Y}
\sim \mathcal{N}_{k+1}\!\left(
(\mathbf{X}^\top \mathbf{X})^{-1}\mathbf{X}^\top\mathbf{Y},
\;\sigma^2(\mathbf{X}^\top \mathbf{X})^{-1}
\right).
\]

\paragraph{Multivariate normal prior with known parameters.}
We also consider a multivariate normal prior with known mean $\boldsymbol{\mu}$ and covariance matrix $\mathbf{C}$,
\[
\boldsymbol{\beta} \sim \mathcal{N}_{k+1}(\boldsymbol{\mu}, \mathbf{C}).
\]
The resulting full conditional distribution is
\[
\boldsymbol{\beta} \mid \sigma^2, \mathbf{X}, \mathbf{Y}
\sim \mathcal{N}_{k+1}\!\left(
(\mathbf{C}^{-1} + \sigma^{-2}\mathbf{X}^\top \mathbf{X})^{-1}
(\mathbf{C}^{-1}\boldsymbol{\mu} + \sigma^{-2}\mathbf{X}^\top \mathbf{Y}),
\;
(\mathbf{C}^{-1} + \sigma^{-2}\mathbf{X}^\top \mathbf{X})^{-1}
\right).
\]

\paragraph{Prior for the error variance $\sigma^2$.}
We assign an inverse-gamma prior to the error variance,
\[
\sigma^2 \sim \text{Inverse-Gamma}(a, b),
\]
with hyperparameters $a,b>0$. Under this prior and the normal likelihood,
the full conditional distribution for $\sigma^2$ is
\[
\sigma^2 \mid \boldsymbol{\beta}, X, Y
\sim \text{Inverse-Gamma}\!\left(
\frac{T}{2}+a,\;
\left(
\frac{1}{2}
\left(
Y^\top Y
- \boldsymbol{\beta}^\top X^\top Y
- Y^\top X \boldsymbol{\beta}
+ \boldsymbol{\beta}^\top X^\top X \boldsymbol{\beta}
\right)
+ \frac{1}{b}
\right)^{-1}
\right).
\]

\section{Evaluation Metrics}
\label{scoringrulesdef}

Predictive accuracy and uncertainty are evaluated using two scoring rules: the modified mean interval score (mMIS) and the continuous ranked probability score (CRPS). The mMIS evaluates the quality of prediction intervals at coverage level $1-\alpha$ by penalizing both wide intervals and intervals that fail to contain the observed value. Let $z_i$ denote the observed response for test observation $i$, and let $[L_i,U_i]$ denote the corresponding $(1-\alpha)$ prediction interval for the $i$th observation in the validation set. The modified interval score ($\mathrm{mIS}_{\alpha, \gamma}$) and $\mathrm{mMIS}$ are
\[
\mathrm{mIS}_{\alpha, \gamma}(L_i, U_i; z_i)
= \gamma(U_i - L_i) + \frac{2}{\alpha}(L_i - z_i)\,\Ind\{z_i < L_i\}
+ \frac{2}{\alpha}(z_i - U_i)\,\Ind\{z_i > U_i\}.
\]
\[\mathrm{mMIS} = \frac{1}{n_{\text{test}}}\sum_i \mathrm{mIS}_{\alpha, \gamma}(L_i, U_i; z_i).
\]
where $\gamma$ is a term that controls the relative importance assigned to interval width. When $\gamma=1$, the metric reduces to the standard interval score \citep{gneiting2007strictly}, which is a strictly proper scoring rule. For $\gamma \neq 1$, the score permits a trade-off between coverage and interval width but is no longer strictly proper.

The CRPS evaluates the full predictive distribution by measuring the squared distance between the predictive CDF $F$ and the empirical CDF of the observation,
\[
\mathrm{CRPS}(F,y)=\int_{-\infty}^{\infty} (F(z)-\mathbf{1}\{z\ge y\})^2 dz.
\]
The CRPS is a strictly proper scoring rule appropriate for evaluating probabilistic predictions \cite{gneiting2007strictly}. Lower values of both scores indicate better predictive performance. 

We also report the average prediction interval widths, empirical coverage probabilities, and root mean squared prediction errors (RMSE).

\section{Application: Basis Functions Generation}
\label{supp:basis}
We generate spatial basis functions by taking the leading $m$ eigenvectors of a pre-specified spatial covariance matrix. 

We estimated the effective range parameter under two fixed values of the Mat\'ern smoothness parameter, $\nu \in \{0.5, 1.5\}$. A random subset of $15{,}000$ locations was selected from the full dataset to ensure computational feasibility. The log-transformed LST observations were then modeled using a spatial process with a Matérn covariance function.

Specifically, we fit separate models for $\nu = 1.5$ and $\nu = 0.5$, allowing the spatial process to capture all coordinate-driven variation by restricting the mean structure to an intercept. This approach isolates the spatial dependence structure and yields estimates of the effective range parameter under different assumptions about smoothness. The resulting estimates were approximately $0.067$ for $\nu=1.5$ and $0.67$ for $\nu=0.5$, indicating markedly different spatial correlation scales depending on the assumed smoothness.

\section{Additional Simulation Results}
\label{app:simulation}

\paragraph{MC Dropout with no aleatoric variance:}
\begin{table}[H]
\centering
\begin{tabular}{rrlllr}
\toprule
   Set &   m & WDR                & DR           & SDR          &   Ntop \\
\midrule
     1 &  25 & (1.0e-10, 1.2e-05) & (0.10, 0.15) & (4.94, 5.50) &   14.6 \\
     1 & 135 & (3.7e-10, 3.2e-06) & (0.10, 0.15) & (5.50, 6.06) &   16.2 \\
     1 & 250 & (1.0e-10, 3.2e-06) & (0.10, 0.15) & (5.50, 6.06) &   15.2 \\
     2 &  25 & (1.0e-10, 2.4e-07) & (0.10, 0.15) & (3.81, 4.38) &   21.4 \\
     2 & 135 & (3.7e-10, 1.2e-05) & (0.10, 0.15) & (5.50, 6.06) &   18.8 \\
     2 & 250 & (4.9e-09, 2.1e-03) & (0.10, 0.15) & (4.94, 6.06) &   16.4 \\
     3 &  25 & (3.7e-10, 2.4e-07) & (0.10, 0.15) & (4.94, 5.50) &   19.4 \\
     3 & 135 & (1.0e-10, 1.2e-05) & (0.10, 0.15) & (5.50, 6.06) &   19.6 \\
     3 & 250 & (3.7e-10, 1.2e-05) & (0.10, 0.15) & (5.50, 6.06) &   13.2 \\
     4 &  25 & (4.9e-09, 1.2e-05) & (0.10, 0.15) & (3.81, 4.38) &   14.4 \\
     4 & 135 & (1.0e-10, 2.1e-03) & (0.10, 0.15) & (4.94, 6.06) &   17.4 \\
     4 & 250 & (3.7e-10, 2.1e-03) & (0.10, 0.15) & (3.81, 5.50) &   20   \\
\bottomrule
\end{tabular}
\caption{\textbf{Top-performing hyperparameter subregions identified via the cubing algorithm.}
For each spatial setting and basis dimension $m \in \{25, 135, 250\}$, the table reports the aggregated hyperparameter subregion formed by combining the top 5 performing cubes across 25 replicates. The four settings correspond to combinations of smoothness $\nu$ and effective range $\rho$: (1) $\nu=0.5$, $\rho=0.3$; (2) $\nu=1.5$, $\rho=0.3$; (3) $\nu=0.5$, $\rho=0.6$; (4) $\nu=1.5$, $\rho=0.6$. WDR denotes the weight decay range, DR the dropout rate range, and SDR the predictive standard deviation multiplier range. The quantity $N_{\text{top}}$ represents the average number of times a cube within the selected subregion appears among the top 10 performing hyperparameter configurations across the 25 replicates. Results are based on mMIS with $\gamma=1$, giving equal importance to interval width and coverage.}
\label{mc_drop_noaleatoric}
\end{table}

The hyperparameter subregions identified in Table~\ref{mc_drop_noaleatoric} exhibit several consistent patterns across spatial settings and basis dimensions. Most notably, the dropout rate range remains fixed at $(0.10, 0.15)$ across all settings and values of $m$, indicating a strong preference for low dropout in achieving optimal interval performance. In contrast, the standard deviation multiplier (SDR) ranges show more variability across settings, with smoother processes (higher $\nu$) and larger effective ranges ($\rho$) generally favoring slightly smaller SDR values, while rougher settings tend to require larger multipliers to maintain adequate coverage.

Weight decay ranges (WDR) vary more substantially, particularly as $m$ increases, suggesting that regularization becomes more sensitive in higher-dimensional basis representations. In several cases, the upper bound of WDR expands by multiple orders of magnitude when moving from $m=25$ to $m=250$, reflecting the need for stronger regularization as model complexity increases.

The values of $N_{\text{top}}$ indicate that the identified subregions appear consistently among the top-performing configurations across replicates, with most values ranging between approximately 14 and 21. Higher $N_{\text{top}}$ values are generally observed for smaller $m$, suggesting that optimal regions are more stable in lower-dimensional settings, while increased variability at larger $m$ reflects a more complex and sensitive hyperparameter landscape. Overall, these results demonstrate that while certain hyperparameters (such as dropout) exhibit robust and consistent behavior, others (notably weight decay and SDR) adapt to both the spatial structure of the data and the dimensionality of the basis representation.

The results in Table~\ref{mc_drop_noaleatoric} reflect the performance of MC Dropout using the cubing-selected hyperparameter subregions without explicitly modeling aleatoric uncertainty. In contrast to the Bayesian regression baseline, uncertainty is controlled primarily through the predictive standard deviation multiplier (SDR), as well as regularization via dropout and weight decay.

In conjunction with the hyperparameter ranges identified in Table~\ref{mc_drop_noaleatoric}, a clear pattern emerges: relatively large SDR values (typically in the range of approximately 4 to 6) are required to achieve near-nominal coverage. This indicates that, in the absence of an explicit variance model, predictive uncertainty must be inflated post hoc to compensate for unmodeled aleatoric variability. Despite this, coverage remains slightly below the nominal 95\% level in several settings, particularly for smaller values of $m$, suggesting that this adjustment is imperfect.

Dropout rates remain consistently low across all settings, reinforcing the observation that excessive stochastic regularization degrades predictive accuracy in this context. Weight decay, however, shows increased variability across both settings and basis dimensions, reflecting its role in stabilizing higher-dimensional representations as $m$ increases.

The high SDR indicates that simply assuming two standard deviations, will not provide optimal coverage and intervals  results demonstrate that while cubing can identify stable and effective hyperparameter regions, the lack of an explicit aleatoric component necessitates larger uncertainty inflation.\newline

\begin{table}[H]
\centering
\begin{tabular}{rrrrrr}
\toprule
   Set &   m &   mMIS &   RMSE &   Width &   Cvg \\
\midrule
     1 &  25 &  2.706 &  0.506 &   1.931 & 0.913 \\
     1 & 135 &  2.004 &  0.391 &   1.557 & 0.936 \\
     1 & 250 &  1.825 &  0.359 &   1.382 & 0.928 \\
     2 &  25 &  1.852 &  0.359 &   1.52  & 0.94  \\
     2 & 135 &  1.317 &  0.254 &   1.057 & 0.935 \\
     2 & 250 &  1.607 &  0.304 &   1.278 & 0.937 \\
     3 &  25 &  1.913 &  0.38  &   1.6   & 0.947 \\
     3 & 135 &  1.769 &  0.35  &   1.356 & 0.929 \\
     3 & 250 &  1.596 &  0.318 &   1.244 & 0.931 \\
     4 &  25 &  1.637 &  0.289 &   1.096 & 0.897 \\
     4 & 135 &  1.052 &  0.196 &   0.852 & 0.941 \\
     4 & 250 &  1.182 &  0.231 &   0.926 & 0.934 \\
\bottomrule
\end{tabular}
\caption{\textbf{Performance of cubing-selected hyperparameter subregions across spatial settings.}
Results are reported across four data-generating settings and basis dimensions $m \in \{25, 135, 250\}$, where $m$ denotes the number of leading eigenvectors used to represent spatial basis functions. The four settings correspond to combinations of smoothness $\nu$ and effective range $\rho$: (1) $\nu=0.5$, $\rho=0.3$; (2) $\nu=1.5$, $\rho=0.3$; (3) $\nu=0.5$, $\rho=0.6$; (4) $\nu=1.5$, $\rho=0.6$. For each setting and $m$, performance is evaluated using mean modified interval score (mMIS), root mean squared error (RMSE), average prediction interval width (Width), and empirical coverage (Cvg). Results are averaged over 25 replicates using hyperparameter subregions identified by the cubing algorithm. The mMIS metric is computed with $\gamma=1$, assigning equal importance to interval width and coverage.}
\label{mc_drop_noaleatoric_result}
\end{table}

The results in Table~\ref{mc_drop_noaleatoric_result} show that MC Dropout with cubing selected hyperparameter subregions achieves competitive predictive performance, though with some systematic differences relative to the Bayesian regression baseline. Across all settings, increasing the number of basis functions $m$ generally leads to improvements in mMIS, RMSE, and interval width, with the largest gains occurring between $m=25$ and $m=135$ and more modest improvements thereafter. This mirrors the trend observed in the Bayesian regression results, indicating that richer basis representations improve both accuracy and uncertainty quantification in both approaches.

However, compared to the Bayesian regression baseline, MC Dropout exhibits a different calibration profile. While interval widths are generally smaller, coverage frequently falls below the nominal 95\% level, particularly for smaller values of $m$ and in settings with smoother underlying processes. This reflects a tendency toward undercoverage, which is consistent with the absence of an explicit aleatoric variance component. In contrast, the Bayesian regression model maintains near-nominal coverage across all settings, albeit with wider intervals.

The trade-off between sharpness and calibration is therefore more pronounced in the MC Dropout baseline. Although lower mMIS values are achieved in several settings, these gains are partially driven by reduced interval width rather than consistently well-calibrated uncertainty. Overall, these results suggest that while cubing is effective in identifying high-performing hyperparameter regions, the lack of explicit modeling of aleatoric uncertainty leads to intervals that are sharper but less reliably calibrated than those produced by Bayesian regression.

\paragraph{MC Dropout with fixed aleatoric variance:}

\begin{table}[H]
\centering
\begin{tabular}{rrlllr}
\toprule
   Set &   m & WDR                & DR           & SDR          &   Ntop \\
\midrule
     1 &  25 & (1.3e-09, 1.8e-08) & (0.10, 0.20) & (2.12, 4.38) &   21.8 \\
     1 & 135 & (4.9e-09, 2.4e-07) & (0.10, 0.30) & (1.00, 3.25) &   21   \\
     1 & 250 & (6.5e-08, 2.4e-07) & (0.15, 0.30) & (1.00, 1.56) &   23.8 \\
     2 &  25 & (1.0e-10, 1.8e-08) & (0.10, 0.15) & (1.56, 4.38) &   24.6 \\
     2 & 135 & (3.7e-10, 1.8e-08) & (0.10, 0.55) & (1.00, 4.38) &   19.4 \\
     2 & 250 & (3.7e-10, 1.8e-08) & (0.10, 0.55) & (1.00, 4.38) &   20.4 \\
     3 &  25 & (6.5e-08, 2.4e-07) & (0.10, 0.25) & (1.56, 2.12) &   25   \\
     3 & 135 & (1.3e-09, 2.4e-07) & (0.10, 0.20) & (1.56, 4.38) &   22.8 \\
     3 & 250 & (1.3e-09, 2.4e-07) & (0.10, 0.20) & (1.56, 4.38) &   22   \\
     4 &  25 & (4.9e-09, 2.4e-07) & (0.10, 0.25) & (1.00, 2.12) &   19.4 \\
     4 & 135 & (4.9e-09, 1.8e-08) & (0.10, 0.35) & (1.56, 2.12) &   18.4 \\
     4 & 250 & (4.9e-09, 2.4e-07) & (0.10, 0.35) & (1.00, 2.12) &   16.2 \\
\bottomrule
\end{tabular}
\caption{\textbf{Top-performing hyperparameter subregions for MC Dropout with fixed aleatoric variance.}
For each spatial setting and basis dimension $m \in \{25, 135, 250\}$, the table reports the aggregated hyperparameter subregion obtained by combining the top 5 performing cubes across 25 replicates using the cubing algorithm. The four settings correspond to combinations of smoothness $\nu$ and effective range $\rho$: (1) $\nu=0.5$, $\rho=0.3$; (2) $\nu=1.5$, $\rho=0.3$; (3) $\nu=0.5$, $\rho=0.6$; (4) $\nu=1.5$, $\rho=0.6$. WDR denotes the weight decay range, DR the dropout rate range, and SDR the predictive standard deviation multiplier range. Predictive uncertainty incorporates a fixed aleatoric component computed using the inverse model precision from the Gal (2016) formulation, combined with MC Dropout epistemic uncertainty. The quantity $N_{\text{top}}$ represents the average number of times a cube within the selected subregion appears among the top 10 performing hyperparameter configurations across the 25 replicates. Results are based on mMIS with $\gamma=1$, assigning equal importance to interval width and coverage.}
\label{mc_drop_fixed}
\end{table}

The hyperparameter subregions identified in Table~\ref{mc_drop_fixed} exhibit several notable differences compared to the baseline MC Dropout results without explicit aleatoric modeling. Most prominently, the SDR ranges are substantially reduced across all settings, with values frequently concentrated between approximately 1 and 3. This contrasts with the larger SDR ranges observed previously, and reflects the incorporation of a fixed aleatoric variance component via the Gal (2016) formulation. As a result, less post hoc inflation of predictive uncertainty is required, since a portion of the variability is now captured directly through the model precision term.

Dropout ranges are also more flexible in this setting, with several configurations allowing moderate dropout values up to 0.30 or higher, particularly for larger $m$. This suggests that once aleatoric uncertainty is accounted for, the model can tolerate greater stochastic regularization without degrading predictive performance. Weight decay ranges remain relatively small but show less extreme variation compared to the no-aleatoric case, indicating a more stable regularization landscape across both settings and basis dimensions.

The values of $N_{\text{top}}$ are generally higher and more consistent across settings, often exceeding 20, which indicates that the identified subregions appear frequently among the top-performing configurations across replicates. This suggests improved stability in the hyperparameter landscape when aleatoric uncertainty is incorporated. Overall, these results demonstrate that introducing a fixed aleatoric component reduces the need for large uncertainty scaling, stabilizes hyperparameter selection, and leads to more coherent and consistent high-performing regions compared to the baseline MC Dropout approach.\newline

\begin{table}[H]
\centering
\begin{tabular}{rrrrrr}
\toprule
   Set &   m &   mMIS &   RMSE &   Width &   Cvg \\
\midrule
     1 &  25 &  2.501 &  0.506 &   1.979 & 0.937 \\
     1 & 135 &  1.876 &  0.396 &   1.502 & 0.941 \\
     1 & 250 &  1.692 &  0.356 &   1.494 & 0.966 \\
     2 &  25 &  1.82  &  0.363 &   1.47  & 0.938 \\
     2 & 135 &  1.235 &  0.26  &   1.05  & 0.95  \\
     2 & 250 &  1.508 &  0.305 &   1.116 & 0.92  \\
     3 &  25 &  1.821 &  0.39  &   1.554 & 0.953 \\
     3 & 135 &  1.694 &  0.351 &   1.375 & 0.943 \\
     3 & 250 &  1.518 &  0.317 &   1.332 & 0.959 \\
     4 &  25 &  1.486 &  0.29  &   1     & 0.904 \\
     4 & 135 &  0.927 &  0.193 &   0.802 & 0.957 \\
     4 & 250 &  1.108 &  0.228 &   0.949 & 0.955 \\
\bottomrule
\end{tabular}
\caption{\textbf{Performance of MC Dropout with fixed aleatoric variance (Gal formulation) across spatial settings.}
Results are reported across four data-generating settings and basis dimensions $m \in \{25, 135, 250\}$, where $m$ denotes the number of leading eigenvectors used to represent spatial basis functions. The four settings correspond to combinations of smoothness $\nu$ and effective range $\rho$: (1) $\nu=0.5$, $\rho=0.3$; (2) $\nu=1.5$, $\rho=0.3$; (3) $\nu=0.5$, $\rho=0.6$; (4) $\nu=1.5$, $\rho=0.6$. For each setting and $m$, performance is evaluated using mean modified interval score (mMIS), root mean squared error (RMSE), average prediction interval width (Width), and empirical coverage (Cvg). Predictive uncertainty combines MC Dropout epistemic uncertainty with a fixed aleatoric component derived from the inverse model precision in the Gal (2016) formulation. Results are averaged over 25 replicates, with mMIS computed using $\gamma=1$ to assign equal importance to interval width and coverage.}
\label{mc_drop_fixed_res}
\end{table}

The results in Table~\ref{mc_drop_fixed_res} demonstrate that incorporating a fixed aleatoric component via the Gal (2016) formulation leads to predictive performance that is broadly comparable to the Bayesian regression baseline, with some notable differences. As in the Bayesian regression results, increasing the number of basis functions $m$ generally improves performance, with reductions in mMIS, RMSE, and interval width, and diminishing gains beyond $m=135$.

In comparison to Bayesian regression, MC Dropout with fixed aleatoric variance typically produces narrower prediction intervals while achieving coverage levels that are closer to the nominal 95\% target than the baseline MC Dropout model without aleatoric modeling. However, coverage remains somewhat more variable across settings, occasionally exceeding or falling below the nominal level, whereas Bayesian regression maintains consistently stable calibration.

The reduction in interval width relative to Bayesian regression suggests that the combined epistemic and fixed aleatoric uncertainty yields sharper predictions, while still maintaining reasonable coverage. This indicates that the Gal-based formulation provides a more balanced trade-off between sharpness and calibration than the no-aleatoric case, though it does not fully replicate the consistency in coverage observed in the Bayesian regression baseline.





\subsection{Algorithim results with CRPS}
\label{alg_res_crps}

\paragraph{Spatial Bayesian Regression Results:}
\label{slr_res_crps}

\begin{table}[H]
\centering
\begin{tabular}{rrr}
\toprule
   Set &   m &   Baseline CRPS \\
\midrule
     1 &  25 &          0.4382 \\
     1 & 135 &          0.3078 \\
     1 & 250 &          0.2753 \\
     2 &  25 &          0.3932 \\
     2 & 135 &          0.1867 \\
     2 & 250 &          0.16   \\
     3 &  25 &          0.3474 \\
     3 & 135 &          0.2389 \\
     3 & 250 &          0.2149 \\
     4 &  25 &          0.2334 \\
     4 & 135 &          0.1307 \\
     4 & 250 &          0.132  \\
\bottomrule
\end{tabular}
\caption{\textbf{Continuous ranked probability score (CRPS) for the Bayesian regression baseline across spatial settings.}
Results are reported across four data-generating settings and basis dimensions $m \in \{25, 135, 250\}$, where $m$ denotes the number of leading eigenvectors used to represent spatial basis functions. The four settings correspond to combinations of smoothness $\nu$ and effective range $\rho$: (1) $\nu=0.5$, $\rho=0.3$; (2) $\nu=1.5$, $\rho=0.3$; (3) $\nu=0.5$, $\rho=0.6$; (4) $\nu=1.5$, $\rho=0.6$. CRPS evaluates the full predictive distribution by jointly assessing calibration and sharpness, with lower values indicating better probabilistic forecasts. Results are averaged over 25 replicates.}
\label{base_crps_br}
\end{table}

The CRPS results in Table~\ref{base_crps_br} show clear and consistent improvements across both spatial settings and basis dimensions. For all settings, CRPS decreases as the number of basis functions $m$ increases, with the largest reductions occurring between $m=25$ and $m=135$, and smaller gains thereafter. This indicates that richer basis representations improve the overall quality of the predictive distribution.

Additionally, CRPS improves as the underlying spatial process becomes smoother and exhibits longer-range dependence. Specifically, moving from Setting 1 ($\nu=0.5$, $\rho=0.3$) to Setting 4 ($\nu=1.5$, $\rho=0.6$), CRPS values decrease substantially across all values of $m$. This suggests that the Bayesian regression model is better able to capture smoother spatial structure with stronger correlation, leading to more accurate and well-calibrated predictive distributions.

\paragraph{MC Dropout with no learned aleatoric variance modeling:}

\begin{table}[H]
\centering
\begin{tabular}{rllrr}
\toprule
   m & WDR                    & DR           &   CRPS &   N\_top \\
\midrule
  25 & (1.00e-10, 1.15e-05) & (0.10, 0.15) & 0.3246 &        17.6 \\
 135 & (3.65e-10, 1.15e-05) & (0.35, 0.45) & 0.2497 &        10.4 \\
 250 & (1.00e-10, 1.77e-08) & (0.45, 0.55) & 0.2219 &        11   \\
  25 & (1.00e-10, 3.16e-06) & (0.10, 0.15) & 0.2206 &        20.2 \\
 135 & (3.65e-10, 1.15e-05) & (0.35, 0.55) & 0.1559 &        11.2 \\
 250 & (1.00e-10, 1.15e-05) & (0.45, 0.55) & 0.1826 &        11.4 \\
  25 & (1.00e-10, 3.16e-06) & (0.10, 0.15) & 0.2428 &        15   \\
 135 & (1.00e-10, 2.37e-07) & (0.45, 0.55) & 0.219  &        10   \\
 250 & (3.65e-10, 3.16e-06) & (0.45, 0.55) & 0.193  &         8   \\
  25 & (1.00e-10, 3.16e-06) & (0.15, 0.25) & 0.1758 &        14.4 \\
 135 & (4.87e-09, 1.15e-05) & (0.45, 0.55) & 0.1172 &        12   \\
 250 & (4.87e-09, 1.54e-04) & (0.45, 0.55) & 0.1363 &         9.6 \\
\bottomrule
\end{tabular}
\caption{\textbf{Top-performing hyperparameter configurations for MC Dropout (no aleatoric variance) evaluated using CRPS.}
Results are reported across four data-generating settings and basis dimensions $m \in \{25, 135, 250\}$, where $m$ denotes the number of leading eigenvectors used to represent spatial basis functions. The four settings correspond to combinations of smoothness $\nu$ and effective range $\rho$: (1) $\nu=0.5$, $\rho=0.3$; (2) $\nu=1.5$, $\rho=0.3$; (3) $\nu=0.5$, $\rho=0.6$; (4) $\nu=1.5$, $\rho=0.6$. WDR denotes the weight decay range and DR the dropout rate range. For each setting and $m$, the table reports the best-performing hyperparameter configurations identified via the cubing procedure. CRPS evaluates the full predictive distribution, with lower values indicating better probabilistic forecasts. The quantity $N_{\text{top}}$ represents the average number of times a configuration appears among the top 10 across 25 replicates.}
\label{no_ale_crps}
\end{table}

This section applies to both the baseline (no explicit modeling of aleatoric variance) and the fixed approach (where aleatoric variance is incorporated via the \citep{gal2016dropout} formulation), as the resulting CRPS values are identical in our implementation. This occurs because CRPS is evaluated using the predictive distributions generated from MC Dropout samples, which are unchanged between the two approaches. While the fixed method augments predictive variance during interval construction, this additional variance is not reflected in the sampled predictive distributions used for CRPS computation. As a result, both approaches In contrast, methods that explicitly learn aleatoric variance, such as the two-headed approach, alter the predictive distribution itself and therefore lead to different CRPS outcomes. 

The hyperparameter configurations in Table~\ref{no_ale_crps} reveal several notable patterns. Weight decay ranges remain consistently very small, often concentrated near $10^{-10}$, though occasional expansions to larger values occur for higher $m$, suggesting increased sensitivity to regularization as model complexity grows. In contrast to earlier mMIS results, the dropout ranges here shift toward substantially higher values for larger $m$ (often between $0.35$ and $0.55$), indicating that stronger stochastic regularization is beneficial when optimizing CRPS, likely to control overfitting in higher-dimensional representations.

In terms of performance, MC Dropout without aleatoric modeling is superior with the Bayesian regression baseline, particularly for smaller values of $m$, where CRPS values are substantially improved. However, for larger $m$, the difference in performance is diminished with the baseline Bayesian regression CRPS values being only slightly higher than MC Dropout.

The values of $N_{\text{top}}$ exhibit greater variability compared to previous tables, with several configurations—especially for larger $m$—appearing relatively infrequently among the top-performing sets. This suggests a less stable hyperparameter landscape under CRPS optimization in the absence of aleatoric modeling, where performance is more sensitive to specific hyperparameter choices. Overall, these results indicate that while higher dropout can partially compensate for the lack of aleatoric variance, the resulting hyperparameter regions are less consistent and more dependent on the specific setting and model dimension.

\paragraph{MC Dropout with learned aleatoric variance modeling:}

\begin{table}[H]
\centering
\begin{tabular}{rllrr}
\toprule
   m & WDR                    & DR           &   CRPS &   N\_top \\
\midrule
  25 & (1.00e-10, 1.155e-05) & (0.10, 0.15) & 0.2724 &        24.8 \\
 135 & (6.49e-08, 1.155e-05) & (0.15, 0.25) & 0.2232 &        12.2 \\
 250 & (4.21e-05, 5.623e-04) & (0.10, 0.25) & 0.2051 &        12.6 \\
  25 & (1.00e-10, 1.155e-05) & (0.10, 0.15) & 0.2013 &        25   \\
 135 & (1.00e-10, 1.540e-04) & (0.10, 0.15) & 0.1442 &        23.6 \\
 250 & (1.00e-10, 1.155e-05) & (0.10, 0.15) & 0.1714 &        21   \\
  25 & (1.00e-10, 1.155e-05) & (0.10, 0.15) & 0.2116 &        25   \\
 135 & (1.00e-10, 1.155e-05) & (0.10, 0.20) & 0.1985 &        12.2 \\
 250 & (1.00e-10, 1.540e-04) & (0.10, 0.15) & 0.18   &        12   \\
  25 & (1.00e-10, 1.540e-04) & (0.10, 0.15) & 0.1624 &        25   \\
 135 & (8.66e-07, 1.540e-04) & (0.10, 0.15) & 0.1093 &        22   \\
 250 & (6.49e-08, 2.054e-03) & (0.10, 0.15) & 0.1295 &        20.6 \\
\bottomrule
\end{tabular}
\caption{\textbf{Top-performing hyperparameter configurations for MC Dropout with learned (heteroskedastic) aleatoric variance evaluated using CRPS.}
Results are reported across four data-generating settings and basis dimensions $m \in \{25, 135, 250\}$, where $m$ denotes the number of leading eigenvectors used to represent spatial basis functions. The four settings correspond to combinations of smoothness $\nu$ and effective range $\rho$: (1) $\nu=0.5$, $\rho=0.3$; (2) $\nu=1.5$, $\rho=0.3$; (3) $\nu=0.5$, $\rho=0.6$; (4) $\nu=1.5$, $\rho=0.6$. WDR denotes the weight decay range and DR the dropout rate range. For each setting and $m$, the table reports the best-performing hyperparameter configurations identified via the cubing procedure. Predictive distributions are obtained using a two-headed neural network trained under a heteroskedastic Gaussian negative log-likelihood, allowing for input-dependent aleatoric variance in addition to MC Dropout epistemic uncertainty. CRPS evaluates the full predictive distribution, with lower values indicating better probabilistic forecasts. The quantity $N_{\text{top}}$ represents the number of times a configuration appears among the top 10 across 25 replicates.}
\label{ler_ale_crps}
\end{table}

The results in Table~\ref{ler_ale_crps} exhibit several clear and consistent patterns across both spatial settings and basis dimensions. First, CRPS decreases as the number of basis functions $m$ increases in most settings, with the largest improvements occurring between $m=25$ and $m=135$, and more modest or occasionally reversed gains beyond $m=135$. This suggests that increasing model capacity improves the quality of the predictive distribution up to a point, after which diminishing returns or mild overfitting may occur.

Across settings, lower CRPS values are generally observed for smoother processes with longer effective ranges, with the best performance achieved in Setting~4 ($\nu=1.5$, $\rho=0.6$). This trend mirrors the Bayesian regression baseline and indicates that the model more effectively captures structure in smoother and more strongly correlated spatial regimes.

The hyperparameter ranges exhibit notable stability, with dropout consistently concentrated between $0.10$ and $0.15$ across nearly all configurations, indicating that only modest stochastic regularization is required when aleatoric variance is learned directly. Weight decay ranges vary more with increasing $m$, particularly at $m=250$, where larger values appear more frequently, reflecting the need for stronger regularization in higher-dimensional settings.

The values of $N_{\text{top}}$ are generally high for $m=25$, often exceeding 24, indicating very stable optimal regions in lower-dimensional settings. However, $N_{\text{top}}$ decreases for larger $m$, particularly at $m=135$ and $m=250$, suggesting increased sensitivity of CRPS performance to hyperparameter choices as model complexity grows. Overall, these results demonstrate that learning aleatoric variance leads to both improved predictive distributions (as measured by CRPS) and more stable hyperparameter behavior, particularly in lower-dimensional regimes.

\paragraph{CRPS Results Summary}
CRPS results reinforce these findings from a distributional perspective. The Bayesian regression model performs strongly, particularly in smoother settings, but is limited by its relatively wide predictive distributions. MC Dropout without aleatoric modeling achieves competitive CRPS values at higher $m$, though performance is less stable and often inferior at lower dimensions, with hyperparameter configurations exhibiting greater variability.

Because CRPS evaluates the full predictive distribution, the baseline and fixed aleatoric approaches yield identical results in our implementation, as the additional variance in the fixed method is applied only during interval construction and does not alter the predictive samples. In contrast, the two-headed model modifies the predictive distribution directly by learning input-dependent variance, resulting in consistently improved CRPS values across settings. In many cases, it outperforms the Bayesian baseline, particularly for moderate values of $m$, indicating more efficient and accurate probabilistic predictions.

While the cubing algorithm is effective in identifying hyperparameter subregions for each of the three modeling approaches to improve performance over baseline configurations, the choice of uncertainty modeling approach plays a more critical role. For the simulated Gaussian data, the two-headed MC Dropout model consistently outperforms both the baseline and fixed approaches across mMIS and CRPS metrics. It achieves sharper intervals with reliable coverage and produces predictive distributions that more accurately reflect the underlying data-generating process.

\section{Application Alternative $\nu$ (1.5)}
\label{alternate_nu}
\begin{table}[H]
\centering
\begin{tabular}{rlrrrrr}
\toprule
   m & Prior   &   MIS &   RMSE &   CRPS &   Width &    Cvg \\
\midrule
  20 & MVN     & 0.362 &  0.038 &  0.026 &   0.349 & 99.767 \\
 135 & MVN     & 0.347 &  0.028 &  0.023 &   0.337 & 99.8   \\
 250 & MVN     & 0.345 &  0.026 &  0.023 &   0.335 & 99.8   \\
\bottomrule
\end{tabular}
\caption{\textbf{Bayesian regression baseline performance for the LST application under Setting 2 ($\nu=1.5$, short-range dependence).}
Results are reported for basis dimensions $m \in \{20, 135, 250\}$, where $m$ denotes the number of leading eigenvectors used to represent spatial basis functions. A multivariate normal (MVN) prior is used for the regression coefficients. Performance is evaluated using mean interval score (MIS), root mean squared error (RMSE), continuous ranked probability score (CRPS), average prediction interval width (Width), and empirical coverage (Cvg, \%). Results are computed on the held-out validation set.}
\label{applied_br_1_5}
\end{table}

\begin{table}[H]
\centering
\begin{tabular}{lrrrrrr}
\toprule
 Model    &   m &   MIS &   CRPS &   RMSE &   Width &   Cvg \\
\midrule
 Baseline &  20 & 0.189 &  0.014 &  0.028 &   0.155 & 0.975 \\
 Baseline & 135 & 0.137 &  0.01  &  0.022 &   0.102 & 0.97  \\
 Baseline & 250 & 0.133 &  0.01  &  0.021 &   0.1   & 0.974 \\
 Fixed    &  20 & 0.971 &  0.017 &  0.032 &   0.958 & 0.991 \\
 Fixed    & 135 & 0.955 &  0.011 &  0.023 &   0.946 & 0.996 \\
 Fixed    & 250 & 0.968 &  0.012 &  0.023 &   0.96  & 0.997 \\
 Two\_head &  20 & 0.151 &  0.014 &  0.029 &   0.134 & 0.991 \\
 Two\_head & 135 & 0.104 &  0.009 &  0.022 &   0.086 & 0.984 \\
 Two\_head & 250 & 0.096 &  0.009 &  0.021 &   0.08  & 0.983 \\
\bottomrule
\end{tabular}
\caption{\textbf{MC Dropout performance for the LST application under Setting 2 ($\nu=1.5$, short-range dependence).}
Results are reported for three MC Dropout variants: Baseline (no explicit modeling of aleatoric variance), Fixed (aleatoric variance incorporated during interval construction using the \citep{gal2016dropout} formulation), and Two\_head (heteroskedastic neural network with learned aleatoric variance). Performance is evaluated across basis dimensions $m \in \{20, 135, 250\}$, where $m$ denotes the number of spatial basis functions. Metrics include mean interval score (MIS), continuous ranked probability score (CRPS), root mean squared error (RMSE), average prediction interval width (Width), and empirical coverage (Cvg). Results are computed on the held-out validation set.}
\label{applied_results_1_5}
\end{table}

Results in Table~\ref{applied_results_1_5} show clear differences in predictive performance across modeling approaches under Setting 2. First, the Bayesian regression baseline exhibits extremely high empirical coverage (approximately $99.8\%$) despite nominal $95\%$ intervals constructed via empirical quantiles. This indicates substantial over-dispersion in the predictive distribution. As a result, intervals are wider than necessary, leading to large coverage and MIS values.

In contrast, MC Dropout baseline models substantially improve interval sharpness, with significantly reduced widths and MIS values while maintaining coverage closer to nominal levels (approximately $97\%$). This suggests that stochastic forward passes better capture localized variability in the LST field, producing more efficient uncertainty quantification. Improvements are also reflected in RMSE and CRPS, both of which decrease relative to the Bayesian regression baseline, indicating gains in both point prediction accuracy and distributional calibration.

The fixed aleatoric approach performs markedly worse in terms of MIS and interval width, with extremely large values across all basis dimensions. This behavior is consistent with the identified hyperparameter subregions, where dropout rates extend as high as $0.55$. Such high dropout induces substantial epistemic variance, and when combined with the fixed aleatoric term from the \citep{gal2016dropout} formulation, results in severe overestimation of total predictive uncertainty. This is further supported by the wide spread in MIS values observed during cubing (e.g., maximum MIS exceeding $3.7$ while minimum MIS $\sim0.12$), although the minimum MIS values remain competitive, indicating that well-performing configurations do exist within the subregion and can still be located with a reduce search, the subregions determined is not as consistent as the other modeling approaches. The corresponding RMSE and CRPS values are also slightly degraded relative to the baseline MC Dropout model, reflecting poorer calibration and predictive accuracy.

The two-head model provides the most balanced performance across all metrics. By explicitly learning heteroskedastic (aleatoric) variance, it produces narrower intervals than the fixed approach while maintaining high coverage (approximately $98\%$–$99\%$). This leads to the lowest MIS values among the MC Dropout variants. Additionally, RMSE and CRPS are consistently minimized, indicating improved point predictions and better-calibrated predictive distributions. In the context of LST, where measurement error and environmental variability are inherently spatially heterogeneous, explicitly modeling aleatoric uncertainty allows the model to adapt uncertainty estimates locally rather than relying on a global correction.

Across all models, increasing the number of basis functions $m$ generally improves performance, with consistent reductions in RMSE and CRPS and modest decreases in MIS and interval width. This reflects the ability of higher-dimensional basis representations to capture finer-scale spatial structure in the LST field.

Overall, these results highlight that while MC Dropout alone provides meaningful improvements over the Bayesian regression baseline, explicitly modeling aleatoric variance through a two-head architecture yields the most reliable and well-calibrated uncertainty estimates in this application.

\end{document}